\documentclass[longauth]{aa}
\usepackage[varg]{txfonts}
\usepackage{natbib}
\bibpunct{(}{)}{;}{a}{}{,}

\usepackage{epsfig,gensymb,siunitx,dcolumn,xspace,textcomp,color}
\usepackage{hyperref}
\usepackage{placeins}
   
\usepackage{subcaption}
\usepackage{graphicx}
\usepackage{array}
\usepackage{tabularx}
\usepackage{float}
\usepackage{wrapfig}
\usepackage{multirow}
\usepackage{fancyhdr}
\usepackage{units}
\usepackage{ragged2e}

\usepackage[acronym]{glossaries}

\usepackage{parnotes}

\usepackage[normalem]{ulem}

\begin{document}
\title{The multi-chord stellar occultation on 2019 October 22 by the trans-Neptunian object (84922) 2003~VS$_2$}
\titlerunning{Stellar occultation by 2003~VS$_2$}
\date{Received date /
Accepted date}

\author{M. Vara-Lubiano\inst{\ref{iaa}}
\and G. Benedetti-Rossi\inst{\ref{lesia},\ref{LIneA},\ref{unesp}}
\and P. Santos-Sanz\inst{\ref{iaa}}
\and J. L. Ortiz\inst{\ref{iaa}}
\and B. Sicardy\inst{\ref{lesia}}
\and M. Popescu\inst{\ref{romacademy},\ref{iac}}
\and N. Morales\inst{\ref{iaa}}
\and F. L. Rommel\inst{\ref{onmctic}, \ref{LIneA}}
\and B. Morgado\inst{\ref{lesia},\ref{LIneA},\ref{onmctic}}
\and C. L. Pereira\inst{\ref{onmctic},\ref{LIneA}}
\and A. Álvarez-Candal\inst{\ref{iaa},\ref{onmctic}}
\and E. Fernández-Valenzuela\inst{\ref{florida}}
\and D. Souami\inst{\ref{lesia},\ref{naxys}}
\and D. Ilic\inst{\ref{univBelgrade},\ref{humboldt}}
\and O. Vince\inst{\ref{astrobsBelgrade}}
\and R. Bachev\inst{\ref{nao}}
\and E. Semkov\inst{\ref{nao}}
\and D. A. Nedelcu\inst{\ref{romacademy}}
\and A. \c{S}onka\inst{\ref{romacademy}}
\and L. Hudin\inst{\ref{l04}}
\and M. Boaca\inst{\ref{l04}}\thanks{Deceased}
\and V. Inceu\inst{\ref{l04}}
\and L. Curelaru\inst{\ref{l13}}
\and R. Gherase\inst{\ref{astrocubul}, \ref{l16}}
\and V. Turcu\inst{\ref{clujnapoca}}
\and D. Moldovan\inst{\ref{clujnapoca}}
\and L. Mircea\inst{\ref{clujnapoca}}
\and M. Predatu\inst{\ref{craiova}}
\and M. Teodorescu\inst{\ref{iss}}
\and L. Stoian\inst{\ref{srpac}}
\and A. Juravle\inst{\ref{srpac}}
\and F. Braga-Ribas\inst{\ref{parana},\ref{lesia},\ref{onmctic},\ref{LIneA}}
\and J. Desmars\inst{\ref{ipsa},\ref{imccecnrs}}
\and R. Duffard\inst{\ref{iaa}}
\and J. Lecacheux\inst{\ref{lesia}}
\and J.I.B. Camargo\inst{\ref{onmctic},\ref{LIneA}}
\and M. Assafin\inst{\ref{valongo},\ref{LIneA}}
\and R. Vieira-Martins\inst{\ref{onmctic},\ref{LIneA},\ref{imccecnrs}}
\and T. Pribulla\inst{\ref{tatranska}}
\and M. Hus\'arik\inst{\ref{tatranska}}
\and P. Sivani\v{c}\inst{\ref{tatranska}}
\and A. Pal\inst{\ref{konkoly}}
\and R. Szakats\inst{\ref{konkoly}}
\and C. Kiss\inst{\ref{konkoly}}
\and J. Alonso-Santiago\inst{\ref{oactinaf}}
\and A. Frasca\inst{\ref{oactinaf}}
\and G. M. Szab\'o\inst{\ref{elte},\ref{mta-elte}}
\and A. Derekas\inst{\ref{elte},\ref{mta-elte}}
\and L. Szigeti\inst{\ref{elte}}
\and M. Drozdz\inst{\ref{mtsuhora}}
\and W. Ogloza\inst{\ref{mtsuhora}}
\and J. Skvar\u{c}\inst{\ref{ing}}
\and F. Ciabattari\inst{\ref{mtagliale}}
\and P. Delincak\inst{\ref{pdlink}}
\and P. Di Marcantonio\inst{\ref{INAF}}
\and G. Iafrate\inst{\ref{INAF}}
\and I. Coretti\inst{\ref{INAF}}
\and V. Baldini\inst{\ref{INAF}}
\and P. Baruffetti\inst{\ref{eaon}}
\and O. Kl\"os\inst{\ref{iotaes}}
\and V. Dumitrescu\inst{\ref{iacustele}}
\and H. Miku\u{z}\inst{\ref{CrniVrh},\ref{ljubljana}}
\and A. Mohar\inst{\ref{darksky}}
}


\institute{Instituto de Astrof\'isica de Andaluc\'ia – Consejo Superior de Investigaciones Cient\'ificas, Glorieta de la Astronom\'ia S/N, E-18008 Granada, Spain\label{iaa}
\and LESIA, Observatoire de Paris, Université PSL, Sorbonne Université, Université de Paris, CNRS, 92190 Meudon, France\label{lesia}
\and Laborat\'orio Interinstitucional de e-Astronomia - LIneA \& INCT do e-Universo, Rua Gal. José Cristino 77, Bairro Imperial de S\~ao Crist\'ov\~ao, Rio de Janeiro (RJ), Brazil\label{LIneA}
\and UNESP - S\~ao Paulo State University, Grupo de Din\^amica Orbital e Planetologia, Guaratinguet\'a, SP, 12516-410, Brazil\label{unesp}
\and Astronomical Institute of the Romanian Academy, 5 Cu\c{t}itul de Argint, 040557 Bucharest, Romania\label{romacademy}
\and Instituto de Astrof\'isica de Canarias (IAC), C/V\'ia L\'actea s/n, 38205 La Laguna, Tenerife, Spain\label{iac}
\and Observat\'orio Nacional/MCTI, Rio de Janeiro (RJ), Brazil, Rua Gal. Jos\'e Cristino 77, Bairro Imperial de S\~ao Crist\'ov\~ao, Rio de Janeiro (RJ), Brazil\label{onmctic}
\and Florida Space Institute, UCF, 12354 Research Parkway, Partnership 1 building, Room 211, Orlado, USA\label{florida}
\and naXys, University of Namur, 8 Rempart de la Vierge, Namur, B-5000, Belgium\label{naxys}
\and Department of astronomy, Faculty of Mathematics, University of Belgrade, Studentski trg 16, 11000 Belgrade, Serbia\label{univBelgrade}
\and Humboldt Research Fellow, Hamburger Sternwarte, Universit{\"a}t Hamburg, Gojenbergsweg 112, 21029 Hamburg, Germany\label{humboldt}
\and Astronomical Observatory, Volgina 7, 11000 Belgrade, Serbia\label{astrobsBelgrade}
\and Institute of Astronomy and National Astronomical Observatory, Bulgarian Academy of Sciences, Sofia, Bulgaria\label{nao}
\and ROASTERR-1 Observatory, L04, Cluj Napoca, Romania\label{l04}
\and Stardust Observatory, L13, Brasov, Romania\label{l13}
\and Astroclubul Bucure\c sti, Romania\label{astrocubul}
\and Stardreams Observatory, L16, V\u{a}lenii de Monte, Romania\label{l16}
\and Romanian Academy - Cluj-Napoca Branch, Astronomical Observatory Cluj-Napoca, Romania\label{clujnapoca}
\and University of Craiova, Craiova, Romania\label{craiova}
\and Institutul de Stiinte Spatiale (ISS), Atomistilor 409, C.P.: MG-23 Magurele 077125, Ilfov, Romania\label{iss}
\and Societatea Rom\^{a}n\u{a} pentru Astronomie Cultural\u{a} (SRPAC), Timisoara, Romania\label{srpac}
\and Federal University of Technology - Paraná (UTFPR / DAFIS), Curitiba, Brazil\label{parana}
\and Institut Polytechnique des Sciences Avanc\'ees IPSA, 63 boulevard de Brandebourg, F-94200 Ivry-sur-Seine, France.\label{ipsa}
\and Institut de M\'ecanique C\'eleste et de Calcul des \'Eph\'em\'erides, IMCCE, Observatoire de Paris, PSL Research University, CNRS, Sorbonne Universit\'es, UPMC Univ Paris 06, Univ. Lille, 77 Av. Denfert-Rochereau, F-75014 Paris, France.\label{imccecnrs}
\and Universidade Federal do Rio de Janeiro, Observat\'orio do Valongo, Rio de Janeiro, Brazil\label{valongo}
\and Astronomical Institute of the Slovak Academy of Sciences, 059 60 Tatranska Lomnica, Slovakia,\label{tatranska}
\and Konkoly Observatory, Research Centre for Astronomy and Earth Sciences, H-1121 Budapest, Konkoly Thege Miklos ut 15-17, Hungary\label{konkoly}
\and Osservatorio Astrofisico di Catania (OACt-INAF), Via S. Sofia 78, 95123 Catania, Italy\label{oactinaf}
\and ELTE E\"otv\"os Lor\'and University, Gothard Astrophysical Observatory, Szombathely, Hungary\label{elte}
\and MTA-ELTE Exoplanet Research Group, 9700 Szombathely, Szent Imre h. u. 112, Hungary\label{mta-elte}
\and Mt. Suhora Observatory, Pedagogical University of Cracow\label{mtsuhora}
\and Isaac Newton Group of Telescopes, Santa Cruz de La Palma, Spain\label{ing}
\and Osservatorio Astronomico di Monte Agliale, Via Cune Motrone, Borgo a Mozzano, Italy\label{mtagliale}
\and PDlink Observatory, Cadca, Slovakia\label{pdlink}
\and INAF – Osservatorio Astronomico di Trieste, Via Tiepolo 11, I-34143 Trieste, Italy\label{INAF}
\and Gruppo Astrofili Massesi, EAON, Massa, Italy\label{eaon}
\and International Occultation Timing Association - European Section (IOTA/ES)\label{iotaes}
\and Ia cu Stele, Bucharest, Romania\label{iacustele}
\and Crni Vrh Observatory, Crni Vrh nad Idrijo, Slovenia\label{CrniVrh}
\and Faculty of Mathematics and Physics, University of Ljubljana, Slovenia\label{ljubljana}
\and Dark Sky Slovenia, Savlje 89, 1000 Ljubljana, Slovenia\label{darksky}
}

\abstract
{Stellar occultations have become one of the best techniques to gather information about the physical properties of trans-Neptunian objects (TNOs), which are critical objects for understanding the origin and evolution of our Solar System.}
{The purpose of this work is to determine with better accuracy the physical characteristics of the TNO (84922) 2003~VS$_2$ through the analysis of the multi-chord stellar occultation on 2019 October 22 and photometric data collected afterward.}
{We predicted, observed, and analyzed the multi-chord stellar occultation of the Gaia DR2 source 3449076721168026624 (m$_v$~=~14.1 mag) by the plutino object 2003~VS$_2$ on 2019 October 22. We performed aperture photometry on the images collected and derived the times when the star disappeared and reappeared from the observing sites that reported a positive detection. We fitted the extremities of such positive chords to an ellipse using a Monte Carlo method. We also carried out photometric observations to derive the rotational light curve amplitude and rotational phase of 2003~VS$_2$ during the stellar occultation. Combining the results and assuming a triaxial shape, we derived the 3D shape of 2003~VS$_2$.}
{Out of the 39 observatories involved in the observational campaign, 12 sites, located in Bulgaria (one), Romania (ten), and Serbia (one), reported a positive detection; this makes it one of the best observed stellar occultations by a TNO so far. Considering the rotational phase of 2003~VS$_2$ during the stellar occultation and the rotational light curve amplitude derived ($\Delta m$~=~\unit[0.264~$\pm$~0.017]{mag}), we obtained a mean area-equivalent diameter of $D_{\mathrm{A_{eq}}}$~=~\unit[545~$\pm$~13]{\si{\kilo\meter}} and a geometric albedo of 0.134~$\pm$~0.010. By combining the rotational light curve information with the stellar occultation results, we derived the best triaxial shape for 2003~VS$_2$, which has semi-axes \textit{a}~=~\unit[339~$\pm$~5]{\si{\kilo\meter}}, \textit{b}~=~\unit[235~$\pm$~6]{\si{\kilo\meter}}, and \textit{c}~=~\unit[226~$\pm$~8]{\si{\kilo\meter}}. The derived aspect angle of 2003~VS$_2$ is $\theta$~=~59\degree~$\pm$~2\degree or its supplementary $\theta$~=~121\degree~$\pm$~2\degree, depending on the north-pole position of the TNO. The spherical-volume equivalent diameter is $D_\mathrm{V_{eq}}$~=~\unit[524~$\pm$~7]{\si{\kilo\meter}}. If we consider large albedo patches on its surface, the semi-major axis of the ellipsoid could be $\sim$~10~km smaller. These results are compatible with the previous ones determined from the single-chord 2013 and four-chord 2014 stellar occultations and with the effective diameter and albedo derived from Herschel and Spitzer data. They evidence that 2003~VS$_2$’s 3D shape is not compatible with a homogeneous triaxial body in hydrostatic equilibrium, but it might be a differentiated body and/or might be sustaining some stress. No secondary features related to rings or material orbiting around 2003~VS$_2$ were detected.}
{}

\keywords{Kuiper belt objects: individual: 2003 VS$_2$ – Methods: observational – Techniques: photometric}

\maketitle

\section{Introduction}
\label{s.introduction}

Trans-Neptunian objects (TNOs) offer a unique opportunity to better understand the origins and the chemical, dynamical, and collisional evolution of the outer Solar System. Possibly dispatched to further distances from the Sun than Neptune's due to gravitational perturbations after their formation \citep{Gomes2005, Levison2008}, their global composition has been virtually unaffected by solar irradiation, keeping it very similar to that of the primitive nebula \citep{Morbidelli2008}.

In the last decade, stellar occultations by TNOs have proved to be one of the best techniques to determine the size and shape of these objects, show features such as satellites \citep{Sickafoose2019} or rings \citep{BragaRibas2014, Ortiz2015, Ortiz2017}, and reveal possible atmospheres \citep{Hubbard1988,SternTrafton2008,Meza2019}. If we combine this technique with light-reflection measurements, we can also derive their geometric albedo. Furthermore, we can calculate their density if we assume hydrostatic equilibrium or if the body is part of a binary system or has a satellite, in which case we can obtain its mass.

Although this technique has been increasingly used in the past few years, it is still challenging to predict and obtain positive results from multi-chord stellar occultations \citep{Ortiz2020}, mainly due to the uncertainties in the orbits of TNOs. The Second Gaia Data Release \citep[Gaia DR2;][]{Gaia2016a, Gaia2016b, Gaia2018} eases the process by providing peerless accurate positions and proper motions of more than one billion sources. However, the large orbital periods of TNOs and the short time span of the observations result in non-negligible uncertainties in their orbital elements, making astrometric observations close to the stellar occultation date indispensable for a successful event prediction.

The object (84922) 2003~VS$_2$ orbits in the 3:2 mean motion resonance (MMR) with Neptune, which makes it a plutino. This TNO presents a double-peaked rotational light curve with an accurately determined rotation period of \unit[7.41753 $\pm$ 0.00001]{\si{\hour}} \citep{SantosSanz2017}. No satellites nor secondary features have been discovered so far around 2003~VS$_2$. Its near-infrared spectrum reveals the presence of exposed water ice \citep{Barkume2008} which, according to \citet{Mommert2012}, might explain the increase in 2003~VS$_2$'s albedo from the typical value of 0.07 for the plutinos. The orbital elements and most relevant physical characteristics of 2003~VS$_2$ can be found in Table \ref{t:vs}.

\begin{table*}
    \centering
    \caption{Orbital elements and physical characteristics of 2003~VS$_2$.}
    \begin{tabular}{*{10}c}
    \hline\hline
         $a$ & $q$ & $e$ & $i$ & $H_V$ & $P$ & $\Delta m$ & $D_{area,eq}$\tablefootmark{a} & $D_{vol,eq}$& $p_V$\tablefootmark{b} \\
         (au)&(au)&&(\degr)&&(\si{\hour})& (mag)&(\si{\kilo\meter})&(\si{\kilo\meter})&\\
         \hline\hline
         39.249 & 36.386 & 0.073 &14.83& 4.14$\pm$0.07 & 7.41753$\pm$0.00001 & 0.141$\pm$0.009 & 553$^{+36}_{-33}$& 548.3$^{+29.5}_{-44.6}$ & 0.131$^{+0.024}_{-0.013}$\\
         \hline
    \end{tabular}
    \tablefoot{Orbital elements of 2003~VS$_2$ from Jet Propulsion Laboratory (JPL) Small-Body Database Browser (\url{https://ssd.jpl.nasa.gov/sbdb.cgi}): $a$, semi-major axis; $q$, perihelion distance; $e$, eccentricity; $i$, inclination. Absolute magnitude $H_V$  from \citet{AlvarezCandal2016} and priv. comm.. Rotation period $P$ is the preferred value from \cite{SantosSanz2017}. Rotational light curve amplitude ($\Delta m$), spherical volume-equivalent diameter ($D_{vol,eq}$), and albedo ($p_\mathrm{V}$) from \cite{BenedettiRossi2019}. Mean area-equivalent diameter ($D_{area,eq}$) calculated from the values given in \citet{BenedettiRossi2019} and taking into account the rotational phase using Eq. \ref{eq:area_eq_diameter}.\\
    \tablefoottext{a}{\cite{Mommert2012} obtained an area-equivalent diameter of  $D~=~523^{+35.1}_{-34.4}$ km} and \tablefoottext{b}{an albedo of $p_\mathrm{V}~=~0.147^{+0.063}_{-0.043}$}.
    }
    \label{t:vs}
\end{table*}

A recent work was published with results from a four-chord and two single-chord stellar occultations by 2003~VS$_2$ in 2013 and 2014 \citep{BenedettiRossi2019}. In that work, they reconstructed the 3D shape of 2003~VS$_2$ by combining the data from the multi-chord stellar occultation and the rotational light curve obtained. The principal semi-axes that provided the best fit had values of \textit{a}~=~\unit[313.8 $\pm$ 7.1]{\si{\kilo\meter}}, \textit{b}~=~\unit[265.5$^{+8.8}_{-9.8}$]{\si{\kilo\meter}}, and \textit{c}~=~\unit[247.3$^{+26.6}_{-43.6}$]{\si{\kilo\meter}}, being these measurements inconsistent with a Jacobi ellipsoid \citep{Chandrasekhar1987}. This solution was highly affected by the rotational light curve amplitude derived in the paper, which was \unit[0.141 $\pm$ 0.009]{mag}, significantly smaller than the previously moderately large ones reported by \citet{Ortiz2006}, \citet{Sheppard2007}, and \citet{Thirouin2010}. In this work we show that such a rotational light curve amplitude cannot be correct based on new data and other reasoning. \citet{BenedettiRossi2019} also reported the presence of a putative secondary feature detected from one of the observing stations but, without further data, they could not discard if it was due to a companion star or instrumental effects.

In this work we present the results of the multi-chord stellar occultation by 2003~VS$_2$ in 2019 October 22. A total of thirty-nine observatories were involved in the campaign, from which 12 positive chords were obtained, a considerable improvement with respect to the previous occultations by this TNO. Combining the collected data from the stellar occultation and ensuing photometric measurements we derived the 3D shape of 2003~VS$_2$.

\section{Observations}
\label{s.observations}

This section presents the observations carried out to improve the prediction of the stellar occultation, the observations of the actual event, and the observations performed shortly afterward to obtain the rotational light curve of 2003~VS$_2$.

\subsection{Occultation predictions}
\label{ss.predictions}

The stellar occultation on 2019 October 22 was singled out from the systematic searches for TNOs occultation candidate stars carried out by the European Research Council (ERC) Lucky Star\footnote{\url{https://lesia.obspm.fr/lucky-star/index.php}} project collaboration \citep{Desmars2015, Desmars2018}. 
The Lucky Star's NIMA\footnote{Numerical Integration of the Motion of an Asteroid.} ephemeris thus gave the initial prediction\footnote{\url{https://lesia.obspm.fr/lucky-star/occ.php?p=32895}} for the 2019 October 22 stellar occultation, and the candidate star was identified in the Gaia DR2 catalog (source ID: 3449076721168026496; UCAC4\footnote{The fourth U.S. Naval Observatory CCD Astrograph Catalog.} identifier 616-023624). Relevant information about the star, such as its coordinates, proper motions, parallax, and their uncertainties, as well as the star's $G$, $B$, $V$, and $K$ magnitudes, can be found in Table \ref{t:star}.

\begin{table*}
\tiny
\centering
\caption{Main information of the occulted star (Gaia DR2 3449076721168026496\tablefootmark{a}; UCAC4 identifier 616-023624\tablefootmark{a}).}
\label{t:star}
\begin{tabular}{*{11}c}
\hline\hline
RA (ICRF\tablefootmark{b}) & errRA	& Dec (ICRF\tablefootmark{b})  & errDec & pmRA             & pmDec             & Plx   &$G$&$B$&$V$&$K$ \\
       & (mas) &           & (mas)  & (mas yr$^{-1}$)  & (mas yr$^{-1}$)   & (mas) &(mag)&(mag)&(mag)&(mag) \\
\hline
05h 30m 38.0442s	& 0.0401	& + 33\degree~ 07\arcmin~ 01\farcs748	& 0.0330	& -0.34$\pm$0.06 & -1.76$\pm$0.05 & 0.2084$\pm$0.0533 & 14.1625  & 15.030 & 14.150 & 10.968\\
\hline
\end{tabular}
\tablefoot{Star Coordinates in RA and Dec propagated to the occultation epoch (2458779.35870) and respective errors (errRA, errDec), proper motion in RA and Dec and respective errors, absolute stellar parallax with error, and magnitude $G$ from Gaia DR2 \citep{Gaia2016a, Gaia2016b, Gaia2018}. Magnitudes $B$, $V$, and $K$ from the Naval Observatory Merged Astrometric Dataset (NOMAD) catalog \citep{Zacharias2004}.\\
\tablefoottext{a}{From Vizier: \url{https://bit.ly/3vUjwkZ}}
\tablefoottext{b}{International Celestial Reference System.}}
\end{table*}

To reduce the uncertainty on 2003~VS$_2$'s orbit and narrow down the predicted location of the shadow path, we performed two observing runs with two different telescopes a few days before the event.

The first observing run was carried out on 2019 October 5, with the charge-coupled device (CCD) Andor ikon-L camera of the 1.5-m telescope at the Sierra Nevada Observatory (OSN) in Granada, Spain. This camera provides a field of view (FOV) of 7\farcm92 $\times$ 7\farcm92, with an image scale of 0\farcs232 pixel$^{-1}$. The 15 images of this set were acquired in 2$\times$2 binning mode, with no filter and an integration time of 400 s. The average seeing was 1\farcs84. Bias and flat-field frames were taken for standard calibration, which was done afterward following the steps in \citet{FernandezValenzuela2016}.

The second run was taken on 2019 October 8, with the IO:O camera of the Liverpool 2-m Telescope at the Roque de los Muchachos Observatory in La Palma, Spain. This instrument has a FOV of 10\arcmin$\times$10\arcmin and an image scale of 0\farcs15 pixel$^{-1}$. The ten images of this set were acquired in 2$\times$2 binning mode, with the Sloan R-filter and 300 s of integration time. The average seeing was 1\farcs2. Bias and flat-field frames were also taken for standard calibration.

With the OSN data, the obtained offsets with respect to the Jet Propulsion Laboratory (JPL) \#30 orbit were ($-$360 $\pm$ 36) mas in right ascension (RA) and (+4 $\pm$ 25) mas in declination (Dec). The Liverpool data yielded offsets of ($-$368 $\pm$ 12) mas in RA and (+24 $\pm$ 11) mas in Dec.

\begin{figure}[htb!]
    \centering
    \includegraphics[width=0.9\columnwidth]{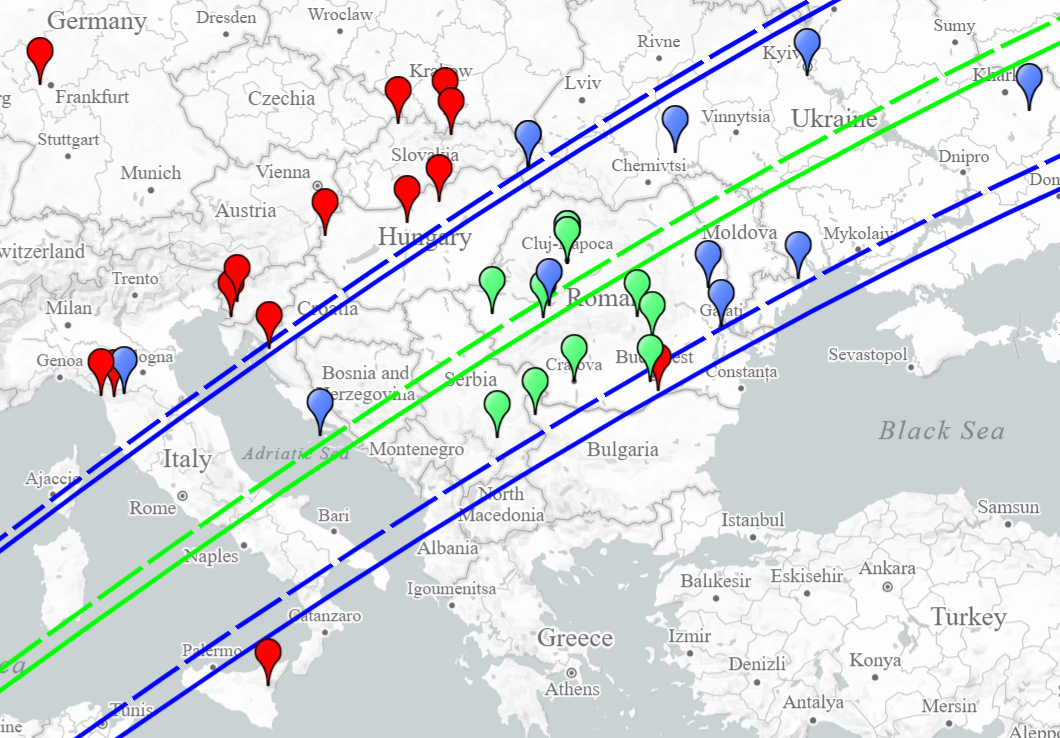}
    \caption{Predicted (solid lines) and observed (dashed lines) shadow paths for the 2003~VS$_2$ stellar occultation on 2019 October 22 through the Gaia DR2 source 3449076721168026496. The prediction was made updating JPL \#30 ephemeris with the offsets obtained with data from the Liverpool 2-m Telescope in the Roque de los Muchachos Observatory (La Palma, Spain). The green line represents the middle of the shadow path and the blue lines indicate the limits of the shadow (width of the shadow path is \unit[479]{\si{\kilo\meter}}, from Jet Propulsion Laboratory). The observing sites involved in the event are also marked: in green, the ones that reported a positive detection; in red, those that reported a negative detection; and, in blue, those that could not observe due to bad weather or technical problems, see Tables \ref{t:pos_sites}, \ref{t:neg_sites}, and \ref{t:noobs_sites}.}
    \label{fig:shadow-path}
\end{figure}

We made two different predictions using the OSN and the Liverpool data, although the result was roughly identical (the Liverpool prediction being $\sim$~20~mas north of the OSN). We used the predicted shadow path obtained by updating 2003~VS$_2$'s ephemeris from JPL with the Liverpool data (shown in Fig. \ref{fig:shadow-path}), as it was taken closer to the event date and with better seeing.

\subsection{Stellar occultation observations}
\label{ss.stellar_occultation}

On 2019 October 22, 39 observational stations spread across 11 countries (including both professional telescopes and amateur observers, Fig. \ref{fig:shadow-path}) were all set to observe the Gaia DR2 source 3449076721168026496 around the predicted occultation time. Out of the total participating stations, 12 reported a positive detection, 14 reported a negative detection (among which two were very close to the shadow path and provide constraints to the shape fitting), and 13 could not observe due to bad weather or technical difficulties. Detailed information about all the participant observatories, divided between positive detections, negative detections, and observations with technical problems or overcast, can be respectively found in Tables \ref{t:pos_sites}, \ref{t:neg_sites}, and \ref{t:noobs_sites}. The detailed instrumental settings at the sites with positive detections are also given in Table \ref{t:pos_sites}. All the teams that could observe collected series of Flexible Image Transport System (FITS) images, except for one site in Romania that took data in Tagged Image File Format (TIFF) format and required a different analysis, as explained in Sect. \ref{ss:time_synch}. The time span of the collected images includes several minutes before and after the stellar occultation.

In this kind of event, it is crucial to save within the images' header the individual acquisition time and all participating stations must be synchronized. Clock synchronizations were used to accomplish it: all the observing teams used Internet Network Time Protocol servers (NTP), except for two negative sites that used Global Positioning System Video Time Inserters (GPS-VTI). Three positive sites had synchronization issues that are addressed in Sect. \ref{ss:time_synch}. 

\begin{figure}
    \resizebox{0.8\hsize}{!}{\includegraphics{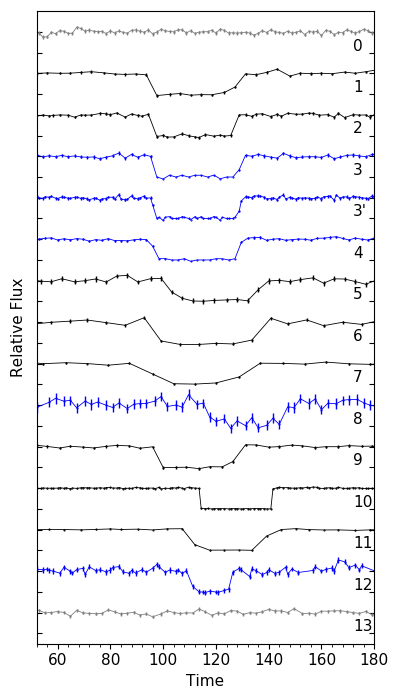}}
    \centering
    \caption{Normalized light curves from the positive detections of the stellar occultation by 2003~VS$_2$ on 2019 October 22 and the two closest negatives. The relative flux of the occulted star with respect to the comparison chosen stars is plotted against time, given in seconds after 2019 October 22 20:40:00 UT. The uncertainty bars of the flux are plotted for all the chords, although some have the size of the points and are not visible. The light curves have been displaced in flux for better visualization and follow the same order as in Table \ref{t:pos_sites}. Chords 0 and 13 (top and bottom chords, in gray) correspond to the negative detections from observers H. Miku\u{z} and V. Dumitrescu, respectively (see Table \ref{t:neg_sites}.) Light curves plotted in blue required a special consideration regarding the time synchronization, see Sect. \ref{ss:time_synch}. Chord 3' is the result of merging chords 3 and 4, see the text for details.}
    \label{fig:normLC}
\end{figure}

We performed synthetic aperture photometry of the occulted star and nearby comparison stars to correct for seeing effects and atmospheric transparency fluctuations. The synthetic aperture photometry was done using our own \texttt{DAOPHOT}-based routines coded in Interactive Data Language (IDL), following the procedures described in \cite{FernandezValenzuela2016}. In this case, though, we fixed the centroid relative to the non-occulted stars in the FOV so that it did not change when the TNO occults the star. We used several apertures and annuli and chose those that resulted in the least scatter in the photometry. In Fig. \ref{fig:normLC} we show the resulting normalized flux (the blended flux of the occulted star + TNO divided by its mean value outside the occultation moment) vs. time from the positive sites, which show deep drops in flux around the predicted occultation time. We note that 2003~VS$_2$ is too faint to be seen with the available equipment, so the star's flux drops to zero during the occultation. We derived the error bars in Fig. \ref{fig:normLC} from Poisson noise calculations, but the results are scaled so that their standard deviation matches that of the data outside the main drop of the occultation (Col. seven in Table \ref{t:occtimes}). We do not see secondary flux drops other than the one corresponding to the main body, which could indicate that there is not a sufficiently wide and dense ring orbiting around 2003~VS$_2$ that could have produced detectable flux drops, but due to the small signal-to-noise ratio (S/N) we cannot discard it. We put special care doing the photometry of the data from the places that missed the occultation but were very close to the shadow path, as they put constraints to 2003~VS$_2$'s final shape.

\begin{table*}[ht]
\centering
\caption{Details of the observing stations of the 2003~VS$_2$ multi-chord stellar occultation on 2019 October 22 with positive detection.}
\label{t:pos_sites}
\begin{tabular}{*6{l}}

        \hline
         Chord  & Site name  & Latitude (N)  & Telescope Aperture    & Exposure Time &  Observers \\
         number & Location   & Longitude (E) & (cm)                  & Cycle Time\tablefootmark{b} &\\
                & MPC Code   & Altitude (m)  & Detector/Instrument   & (s)           &      \\
        \hline\hline

        &ROASTERR-1 Observatory (1) & 46\degr~49\arcmin~15\farcs6 &  30       & 3         &           \\
        1&Romania                   & 23\degr~35\arcmin~47\arcsec   & KAF8300   & 5.18   & L. Hudin  \\
        &L04                        & 391\tablefootmark{a}          &           &           &           \\ \hline

        &ROASTERR-1 Observatory (2)   & 46\degr ~49\arcmin ~15\farcs6     & 28              & 2              &                 \\
        2&Romania    & 23\degr ~35\arcmin ~47\arcsec    & ATIK-414ex & 3.01     & M. Boaca     \\
        &L04        & 391\tablefootmark{a}           &                     &               &  V. Inceu            \\ \hline

        &    & 46\degr~ 42\arcmin~ 37\farcs557   & 30.5               & 2              & V. Turcu      \\
        3&Romania    & 23\degr~ 35\arcmin~ 35\farcs647   & SBIG STT-1603ME     & 3              & D. Moldovan    \\
        &--          & 783.405       &                     &               & L. Mircea    \\ \hline

        &    & 46\degr~ 42\arcmin~ 37\farcs452   & 40.6               & 2              & V. Turcu      \\
        4&Romania    & 23\degr~ 35\arcmin~ 36\farcs94  & SBIG STT-1603ME     & 3              & D. Moldovan    \\
        &--          & 787.56        &                     &               & L. Mircea    \\ \hline

        &         & 45\degr~ 42\arcmin~ 11\farcs94     & 27.94             & 5             & L. Stoian       \\
        5&Romania    & 21\degr~ 26\arcmin~ 12\farcs78     & ZWO ASI 224 CMOS    & 5.14      & A. Juravle      \\
        &--          & 92\tablefootmark{a}           &                     &               &                 \\ \hline

        &Berthelot Observatory & 45\degr~ 36\arcmin~ 59\farcs4 & 36.83             & 2.5              & D. A. Nedelcu        \\
        6&Romania    & 22\degr~ 53\arcmin~ 19\farcs7     & SBIG                & 8.64              & A. Sonka    \\
        &L54          & 400           & STL11000M CCD       &               &                 \\ \hline

        &Stardust Observatory    & 45\degr~ 38\arcmin~30\arcsec    & 20.32             & 8.5           &                 \\
        7&Romania    & 25\degr~ 37\arcmin~19\arcsec    & CCD Atik 383L       & 10.28      & L. Curelaru         \\
        &L13        & 597           &                     &               &         \\ \hline

        &           & 44\degr~ 19\arcmin~ 24\farcs5    & 35.6               & 3.01          &                 \\
        8&Romania    & 23\degr~ 47\arcmin~ 19\farcs0    & ASI 1600            & 3.34          & M. Predatu          \\
        &--          & 109\tablefootmark{a}          &                     &                   &          \\ \hline

        &Stardreams Observatory    & 45\degr~ 12\arcmin~ 13\farcs3    & 20   & 4             &                 \\
        9&Romania    & 26\degr~ 02\arcmin~ 44\farcs2   & ATIK 460ex    & 5.52       & R. Gherase    \\
        &L16        & 382\tablefootmark{a}         &             &               &                 \\ \hline
        
        &Astronomical Station Vidojevica & 43\degr~ 08\arcmin~24\farcs6 & 140          & 0.8              & D. Ilic    \\
        10&Serbia     & 21\degr~ 33\arcmin~20\farcs4  &  Andor ikonL        & 1.55              & O. Vince    \\
        &C89          & 1150          &                     &               &                 \\ \hline

        &Belogradchik Observatory  & 43\degr~ 37\arcmin~22\arcsec & 60      & 6              & R. Bachev          \\
        11&Bulgaria   & 22\degr~ 40\arcmin~30\arcsec   & FLI PL9000          & 6.86              & E. Semkov \\
        &--       & 650          &                     &               &                 \\ \hline

        &    & 44\degr~ 19\arcmin~19\farcs14 & 35.5               & 2              &             \\
        12&Romania    & 25\degr~ 59\arcmin~8\farcs69  & ASI 1600            & 2.50      & M. Teodorescu      \\
        &--          & 70\tablefootmark{a}           &                     &               &                 \\ \hline

\end{tabular}
\tablefoot{The sites are sorted by their distance to the center of the predicted shadow path.\\
\tablefoottext{a}{Altitudes from Google Earth.}
\tablefoottext{b}{Cycle time is the sum of exposure time plus dead time.}}
\end{table*}

\begin{table*}[ht]
    \centering
    \caption{Details of the observing stations of the 2003~VS$_2$ multi-chord stellar occultation on 2019 October 22 with negative detection.}
    \begin{tabular}{*4{l}}
        \hline
        \multicolumn{1}{l}{Site name}  & Latitude (N)  & Telescope & Observers          \\
        \multicolumn{1}{l}{Location} & Longitude (E) & aperture &  \\
        \multicolumn{1}{l}{MPC Code}   & Altitude (m)  & (\si{\centi\meter})                   &           \\ \hline \hline 
        
        Bremthal & 50\degr~ 08\arcmin~ 17\farcs4 & &\\
        Germany & 08\degr~ 21\arcmin~ 50\farcs4  & 25.4 & O. Kl\"os\\
        -- & 256  &&\\\hline
        
        PDlink Observatory & 49\degr~ 24\arcmin~ 15\farcs20 &  &\\
        Cadca, Slovakia & 18\degr~ 42\arcmin~ 09\farcs47  & 40 & P. Delincak\\
        -- & 680 &&\\\hline
        
        Mt. Suhora Observatory & 49\degr~ 34\arcmin~09\arcsec & & M. Drozdz\\
        Poland & 20\degr~ 04\arcmin~03\arcsec  & 60 & W. Ogloza\\
        -- & 1009  &&\\\hline
        
        Skalnate Pleso & 49\degr~ 11\arcmin~21\farcs8 & & T. Pribulla\\
        Slovakia & 20\degr~ 14\arcmin~02\farcs1 & 130 & M. Hus\'arik\\
        056 & 1786 && P. Sivani\v{c}\\\hline

        & 44\degr~ 01\arcmin~16\farcs96 & &\\
        Massa, Italy & 10\degr~ 07\arcmin~56\farcs65  & 30 &P. Baruffetti\\
        -- & 30 &&\\\hline
        
        Gothard Observatory & 47\degr~ 15\arcmin~29\farcs83 & & G. M. Szab\'o\\
        Szombathely, Hungary & 16\degr~ 36\arcmin~15\farcs67 & 80 & A. Derekas\\
        -- & 220\tablefootmark{a} &&L. Szigeti\\\hline
        
        \u{C}rni Vrh Observatory & 45\degr~ 56\arcmin~45\farcs1 &  &\\
        Slovenia & 14\degr~ 04\arcmin~16\farcs4 &  60 & J. Skvar\u{c}\\
        -- & 726  &&\\\hline
        
        Monte Agliale Observatory & 43\degr~ 59\arcmin~ 43\farcs0 & &\\
        Borgo a Mozzano, Italy & 10\degr~ 30\arcmin~ 53\farcs7 & 50 & F. Ciabattari\\
        159 & 756 &  &\\\hline
        
        Stazione Osservativa di Basovizza & 45\degr~ 38\arcmin~ 33\arcsec  &&P. Di Marcantonio\\
        Osservatorio Astronomico di Trieste, Italy & 13\degr~ 52\arcmin~ 23\arcsec  &  35  &G. Iafrate\\
        038 & 400  &&I. Coretti, V. Baldini\\\hline
        
        Piszk\'estető Station & 47\degr~ 55\arcmin~6\farcs0 & & A. Pal\\
        Konkoly Obs., Hungary & 19\degr~ 53\arcmin~41\farcs7 & 100 & R. Szakats\\
        561 				& 924\tablefootmark{a}				& &C. Kiss\\\hline
        
        Konkoly Observatory & 47\degr~ 29\arcmin~59\farcs28 & & A. Pal\\
        Normafa, Hungary    & 18\degr~ 57\arcmin~51\farcs12 & 60 & R. Szakats\\
        053                 & 468\tablefootmark{a}        &&C. Kiss\\\hline
        
         & 49\degr~ 24\arcmin~ 15\farcs20 &  & \\
        Vratnik, Croatia & 18\degr~ 42\arcmin~ 09\farcs47 & 20 &H. Miku\u{z}\\
        -- & 775 && A. Mohar\\\hline

         & 44\degr~ 07\arcmin~56\farcs8  &   &\\
        Romania & 26\degr~ 13\arcmin~04\farcs3 & 20  & V. Dumitrescu\\
        -- & 88\tablefootmark{a} &&\\\hline
  
        Mount Etna Observatory & 37\degr~ 41\arcmin~05\arcsec & & J. Alonso-Santiago\\
        Osservatorio Astrofisico di Catania, Italy & 14\degr~ 58\arcmin~04\arcsec & 91  & A. Frasca\\
        156 & 1735 &&\\\hline
    
    \end{tabular}
    \label{t:neg_sites}
\tablefoot{The sites are sorted by their distance to the center of the predicted shadow path.\\
\tablefoottext{a}{Altitudes from Google Earth.}}
\end{table*}

\subsection{Rotational light curve observations}
\label{ss.rotLCobs}

Given that 2003~VS$_2$ has a double-peaked rotational light curve and is large enough to allow a hydrostatic equilibrium shape \citep[see a detailed explanation in][]{BenedettiRossi2019}, we can assume that the body is a triaxial ellipsoid \citep{Chandrasekhar1987}. If this were the case, one would need at least three body projections at different rotational phases to correctly derive the actual 3D shape of 2003~VS$_2$.

However, we can overcome this by combining the occultation information with time-series photometric data, taken closer to the occultation date. To do so, we observed 2003~VS$_2$ on two consecutive nights, two days after the event (2019 October 24 and 25), with the 1.23-m telescope at the Calar Alto Observatory in Almer\'ia (Spain). This telescope is equipped with a 4k x 4k CCD DLR-MKIII camera, which provides a FOV of 21\farcm5$\times$21\farcm5 with an image scale of 0\farcs32 pixel$^{-1}$. This wide FOV allowed us to keep the same stellar field both nights, making it possible to choose the same reference stars during the run to minimize systematic photometric errors.

A total of 143 images were acquired in 2$\times$2 binning mode and with an integration time of 400\si{\second}, using no filter to maximize the S/N. The moonshine was 14\% for the first night and 7\% for the second night. The average seeing at Calar Alto from the Differential Image Motion monitor was 1\farcs6 for the second night; unfortunately, the Calar Alto seeing tracker did not work for the first night. The average measured full width at half maximum (FWHM) was $\sim$~2.5 pixels ($\sim$~1\farcs6) for the first night and $\sim$~2.3 pixels ($\sim$~1\farcs5) for the second. The S/N was between 30 and 70 during the first night and between 20 and 80 during the second night. We also took bias and flat-field frames each night for standard calibration, which was performed using our own specific routines written in IDL.

\section{Data reduction}

\subsection{Time synchronization and extraction}
\label{ss:time_synch}
Robust clock synchronization is essential to have the absolute acquisition time written on each frame header to obtain the actual star dis- and re-appearance times from each site and, hence, the projected chords' relative position. In this regard, some of the images collected required a different treatment to obtain their absolute acquisition time:

\begin{itemize}
    \item Chords 3 and 4: there was an intentionally applied \unit[1]{\si{\second}} difference between the images acquired from the two telescopes at this site, aiming to cover the whole event avoiding the dead time of the CCDs. Since the two telescopes used the same instrument and took FITS images with the same exposure time, we merged the data from these two telescopes to obtain one single chord. By doing this, we doubled the sampling of the light curve and, therefore, we obtained the immersion and emersion times from this site with a smaller uncertainty. Hereafter we will refer to this as a single chord (chord 3').
    
    \item Chord 8: although this chord was nearly overlapping chord 9, its center was shifted by more than \unit[12]{\si{\second}} with respect to the linear fit of the rest of the centers (see Fig. \ref{fig:normLC} and Table \ref{t:occtimes}). Later tests performed a month after the occultation event showed a difference of more than one minute between the acquisition system and a synchronized clock, but we were not able to find out the exact time offset during the occultation event. All of this suggested that the time synchronization at this site was not correctly applied, so we could only use the chord length but not the absolute time of the chord.
    
    \item Chord 12: the images from this site were recorded in a data cube in TIFF format. We converted them to FITS format using the Planetary Imaging PreProcessor (PIPP), but then all the images had the same time written on the header. The observer provided us with the acquisition times of the images, but their accuracy only reached the order of seconds (as this is the accuracy reached by the software used to save the images). We used the mid-integration times for the light curve and then increased the uncertainties of the immersion and emersion times by the exposure time (two seconds) since it was not possible to know if the acquisition times we have corresponded to the beginning, middle, or end of the images' integration.
    
\end{itemize}

We saw no evidence for wrong time synchronization for the remaining positive chords, given their relative positions (see Fig. \ref{fig:limbfit}a).

\subsection{Occultation times and chord lengths}
\label{ss:square_well_model_fit}

From every site that reported a positive detection, the stellar occultation starting and ending times (disappearance and reappearance of the star, respectively) were determined by fitting each light curve to a sharp-edge square-well model convolved by: (1) Fresnel diffraction by a point-like source at the distance of 2003~VS$_2$ from the observer, (2) the CCD bandwidth, (3) the finite stellar diameter projected at the object's distance, in kilometers, and (4) the finite integration time \citep[see, e.g.,][]{Widemann2009, BragaRibas2013}. 

During the stellar occultation, 2003~VS$_2$ was at a geocentric distance of $\Delta$~=~\unit[36.1]{au}=\unit[5.40$\times10^9$]{\si{\kilo\meter}}; this implies that for a typical wavelength of $\lambda $=0.65 \si{\micro\meter}, the Fresnel scale F$= \sqrt{\lambda D/2}$ has a value of 1.33 \si{\kilo\meter}.

We estimated the projected stellar diameter using the formulae from \citet{vanBelle1999} and its $B$, $V$, and $K$ apparent magnitudes, obtained from the NOMAD catalog \citep[][Table \ref{t:star}]{Zacharias2004}. The diameters obtained are \unit[0.86]{\si{\kilo\meter}} if considering a Super Giant or \unit[0.91]{\si{\kilo\meter}} if considering a Main Sequence star. 

Finally, we converted the shortest integration time among the positive observations (\unit[0.8]{\si{\second}}, chord 10) into the distance in the sky-plane traveled by 2003~VS$_2$ between two adjacent data points; since the velocity of 2003~VS$_2$'s shadow-path was \unit[13.01]{\si{\kilo\meter\per\second}}, that distance was \unit[10.41]{\si{\kilo\meter}}. As a result, our light curves are dominated mainly by the integration times and not by Fresnel diffraction nor the stellar diameter.

\begin{figure}
    \resizebox{\hsize}{!}{\includegraphics{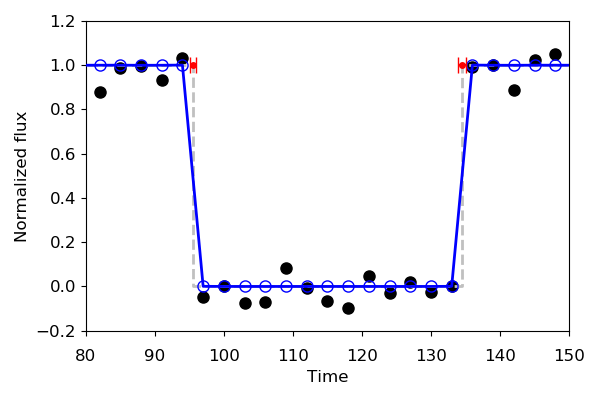}}
    \caption{Best fit of the data from ROASTERR-1 Obs.(2) to the convolved square-well model. The flux from the occulted star + 2003~VS$_2$ is plotted with black dots, normalized to the flux of the star while unocculted. Time is in seconds after 2019 October 22 - 20:40:00 UT. The dashed gray line represents the initial square-well model; the solid blue line represents the final fit, after convolving the square-well model by Fresnel diffraction, the exposure time, and the stellar diameter, see Sect. \ref{ss:square_well_model_fit} for details. The open blue circles show the expected flux from such model. The stellar occultation starting and ending times derived are plotted in red, with their uncertainties. For all the positive chords, the star dis- and re-appearance times and chord lengths derived are listed in Table \ref{t:occtimes}.}
    \label{fig:square-well_fit}
\end{figure}

The fitting procedure then searches for the times of disappearance and reappearance of the star that minimize a classical $\chi ^2$ function, as explained in \citet[suppl. inf.]{Sicardy2011}. The uncertainty bars are estimated by varying the occultation times to increase $\chi ^2$ to $\chi ^2 +1$. Since these uncertainty bars depend on the fit of the photometry and the typical errors in the photometry are very small, the uncertainty in the derived time of ingress and egress is also small, a small fraction of the integration time. Figure \ref{fig:square-well_fit} shows an example of the best fit to one of the stellar occultation light curves (chord 2). All the derived disappearance and reappearance times and chord lengths are listed in Table \ref{t:occtimes}.

\begin{table*}[ht]
\centering
\caption{Star dis- and re-appearance UT times on 2019 October 22, chord lengths, time shifts, and dispersion ($\sigma$) of the light curves outside the occultation.}
\label{t:occtimes}
\begin{tabular}{lccllll}
\hline
Chord number & Disappearance (s)        & Reappearance (s)          & Chord size (s)& Chord Size (km)       & Shift (s)\tablefootmark{a}    & $\sigma$\\\hline\hline
1            & 20:41:34.800 $\pm$ 1.155	& 20:42:14.995 $\pm$ 0.225  &40.2 $\pm$ 1.2 & 523 $\pm$ 16          & +0.7  & 0.06\\
2		     & 20:41:35.54 $\pm$ 0.46	& 20:42:14.47 $\pm$ 0.55    &38.9 $\pm$ 0.7& 506 $\pm$ 9          & +0.6  & 0.06\\
3'\tablefootmark{b} & 20:41:35.609 $\pm$ 0.109  & 20:42:16.585 $\pm$ 0.077  & 40.98 $\pm$ 0.13 & 533.1 $\pm$ 1.7   & +0.03 & 0.06\\
5	        & 20:41:44.375 $\pm$ 0.525	& 20:42:24.95 $\pm$ 0.40    &40.6 $\pm$ 0.7& 528 $\pm$ 9          & -0.06 & 0.11\\
6			& 20:41:37.325 $\pm$ 1.100	& 20:42:24.062 $\pm$ 1.287  &46.7 $\pm$ 1.7& 608 $\pm$ 22          & +1.6  & 0.08\\
7			& 20:41:35.420 $\pm$ 0.425	& 20:42:17.411 $\pm$ 0.382  &42.0 $\pm$ 0.6& 546 $\pm$ 8          & +0.07 & 0.05\\
8           & 20:42:02.05 $\pm$ 0.73    & 20:42:37.76 $\pm$ 0.76    & 35.7$\pm$ 1.1& 464 $\pm$ 14          & -14   & 0.23\\
9			& 20:41:37.491 $\pm$ 0.590	& 20:42:14.897 $\pm$ 0.160  &37.4 $\pm$ 0.6 & 487 $\pm$ 8          & +1.1  & 0.06\\
10			& 20:41:57.914 $\pm$ 0.110	& 20:42:31.944 $\pm$ 0.136  &34.03 $\pm$ 0.17& 443 $\pm$ 2           & +0.3  & 0.02\\
11		    & 20:41:54.60 $\pm$ 0.15	& 20:42:28.76 $\pm$ 0.12    &34.16 $\pm$ 0.19& 444 $\pm$ 2           & -0.7  & 0.02\\
12          & 20:41:53.551 $\pm$ 2.186\tablefootmark{c}	& 20:42:12.949 $\pm$ 2.387\tablefootmark{c}  &19 $\pm$ 3& 247 $\pm$ 39  & -2    & 0.12\\

\hline
\end{tabular}
\tablefoot{Chords' shifts are explained in Sect. \ref{ss:limb_fitting}.\\
\tablefoottext{a}{A positive shift means a shift toward the east, a negative one indicates a shift toward the west.}
\tablefoottext{b}{Chord 3' is the result of merging chords 3 and 4 in Table \ref{t:pos_sites}, see Sect. \ref{ss:time_synch} for details.}
\tablefoottext{c}{We have added two seconds to the uncertainty bar given by the square-well fit for chord 12, as explained in Sect. \ref{ss:time_synch}.}}
\end{table*}

\subsection{Rotational light curve}
\label{ss:rot_LC_analysis}

We performed aperture photometry on 2003~VS$_2$ and 16 reference stars with good photometric behavior using the routines and procedures mentioned in \cite{FernandezValenzuela2016}. We tested different synthetic aperture radii and sky annuli to maximize the S/N of the object while minimizing the dispersion of the residuals to the rotational light curve fit. We chose different aperture parameters for each observing day but selecting the same reference stars and then combined the best photometry results. 

\begin{figure}
    \resizebox{\hsize}{!}{\includegraphics{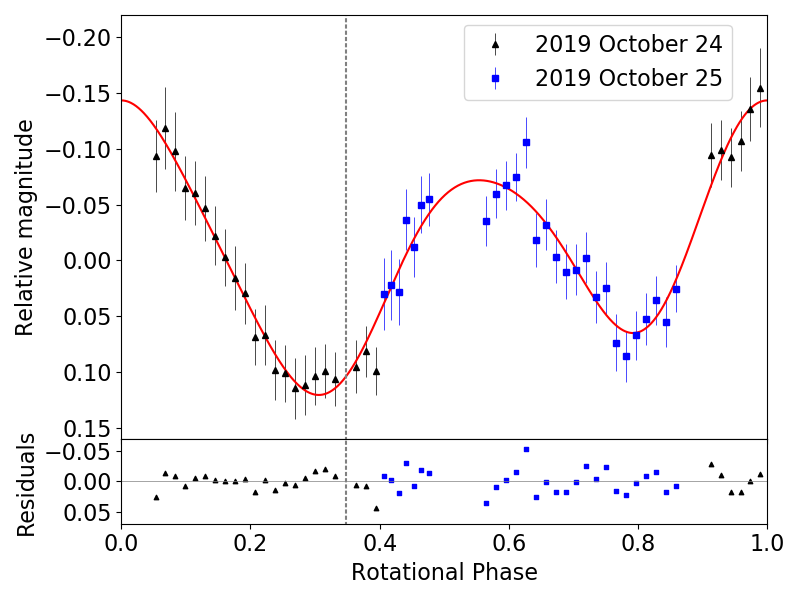}}
    \caption{On top, rotational light curve of 2003~VS$_2$ from data collected in 2019 October 24 (black triangles) and 25 (blue squares). The data were folded using the rotation period of \unit[7.41753]{\si{\hour}} \citep{SantosSanz2017}. The fourth-order Fourier fit is shown in a solid red line. The vertical black-dashed line indicates the rotational phase of 2003~VS$_2$ at the time of the stellar occultation. The plot has been arbitrarily shifted to make the minimum of the fit (maximum brightness) correspond to rotational phase 0. On the bottom, differences between the observational data and the fit. Julian dates are not corrected from light travel time.}
    \label{fig:rotationalLC}
\end{figure}

The obtained rotational light curve is shown in Fig. \ref{fig:rotationalLC} and the data used is available online. We folded the relative flux of 2003~VS$_2$ vs. time using its well-known rotation period of \unit[7.41753 $\pm$ 0.00001]{\si{\hour}} \citep{SantosSanz2017} and used a fourth-order Fourier function to fit the folded data. The obtained peak-to-peak amplitude of the rotational light curve was $\Delta m$~=~\unit[0.264~$\pm$~0.017]{mag}; the nominal value is the one that best fits the data in terms of minimization of the sum of squared residuals and the uncertainty is given as the standard deviation of a Monte Carlo distribution. This amplitude is larger than other values found in the literature \citep[0.23 $\pm$ 0.07 mag, 0.21 $\pm$ 0.02 mag, 0.21 $\pm$ 0.03 mag;][respectively]{Ortiz2006,Sheppard2007,Thirouin2010}, although it is consistent within their error bars, but the greatest disagreement appears when comparing this work's amplitude with the last 0.141 $\pm$ 0.009 mag value found in \citet{BenedettiRossi2019}, which was obtained with data from 2014. Our confidence is stronger for this work's rotational light curve since the dispersion of the residuals to the fit is lower than the one in \citet{BenedettiRossi2019} and the obtained amplitude is more compatible with the ones previously published in the literature. The slight increase in the rotational light curve amplitude along the years may be explained by a change in the object's aspect angle, as found in the case of the TNO Varuna \citep{Fernandez-Valenzuela2019}.

We calculated the rotational phase of 2003~VS$_2$ at the time of the stellar occultation\footnote{To do this, we considered 2458779.36248264 to be the Julian date of the occultation event, as it was the closest recorded date to the average value of the mid-occultation times from all sites. However, choosing a different Julian date close to the stellar occultation would not change the results, given the object's rotation period.}, which was 0.32 with respect to the absolute brightness maximum (see the vertical black dashed line in Fig. \ref{fig:rotationalLC}). We did not correct the Julian dates from light travel time since this source of error is negligible due to the closeness in time of all data. The rotational phase obtained implies that the apparent surface area of 2003~VS$_2$ was near its minimum during the occultation event. We note that the rotational phase at the time of the occultation is not influenced by the amplitude of the rotational light curve but by the rotation period. Given the precision of the rotation period and the time-span between the data taken for the occultation event and the rotational light curve, the error calculating the rotational phase is negligible.

\section{Results analysis}

\subsection{Limb fitting}
\label{ss:limb_fitting}

The generally considered shape for a TNO's limb fitting is an ellipse \citep[see, e.g.,][]{Ortiz2017, BenedettiRossi2019}, even though the object might not project a perfect regular elliptical shape. In this work we do not have enough positive chords to account for topographic features or deviations from a true elliptical shape, so we can only use the simplest and most general model for the limb fitting.

Five adjustable parameters characterize the considered ellipse: the body's center coordinates relative to the star in the sky plane ($f_\mathrm{c}$, $g_\mathrm{c}$), the apparent semi-major axis \textit{a'}, the apparent semi-minor axis \textit{b'}, and the tilt angle of the ellipse \textit{PA}, which is the position angle of the semi-minor axis from celestial north, positive to the west. If we assume the Gaia DR2 star position to be correct, the coordinates ($f_\mathrm{c}$, $g_\mathrm{c}$) respectively give the offsets in RA and Dec to apply to the object's adopted ephemeris.

\begin{figure*}[t!]
    \centering
    \includegraphics[width=17cm]{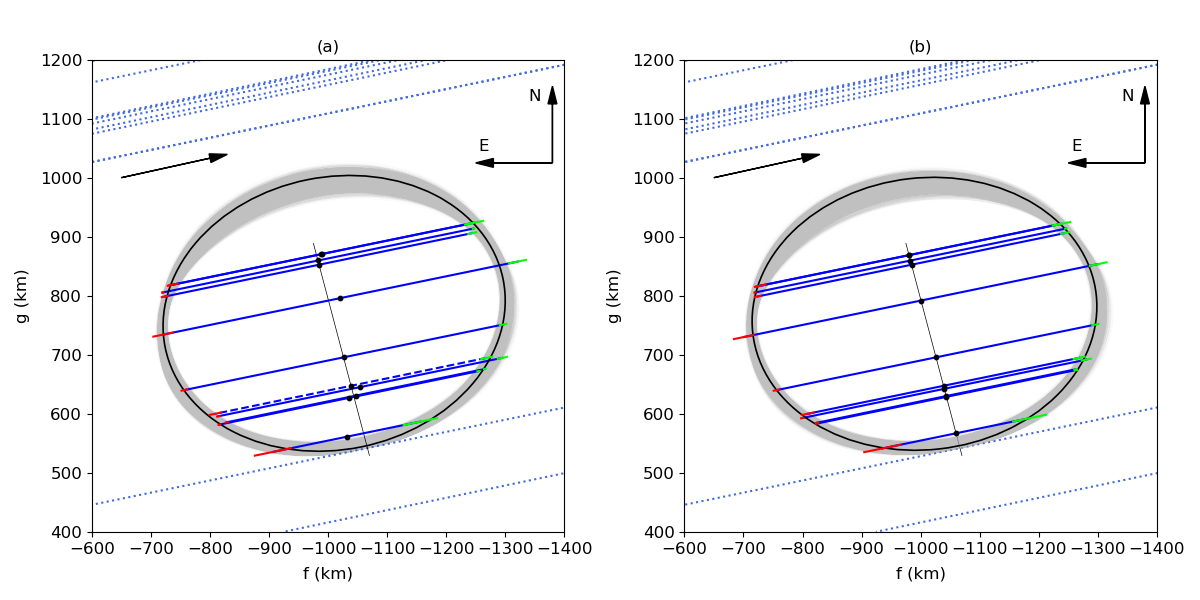}
    \caption{Elliptical fit to the chords of the stellar occultation, for the two considered configurations: (a) original distribution of the chords, with chord 8 (dashed line) already shifted, and (b) final distribution of the chords after aligning their centers using a least squares linear fit. In both plots: the positive chords are shown in solid blue, negative chords in dotted blue; uncertainties of the star's disappearance time are shown in green and those of the reappearance time in red; the black dots show the center of the chords. From top to bottom, the chords follow the same order as in Table \ref{t:pos_sites}. The limiting negative chord on the north corresponds to observer H. Miku\u{z} and the chord limiting on the south corresponds to observer V. Dumitrescu, see Table \ref{t:neg_sites}. The black arrows on the top-left of each plot show the direction of the shadow motion. The best elliptical fit to the extremities of the chords is shown in black and all the ellipses from the Monte Carlo distribution are plotted in gray, see Sect. \ref{ss:limb_fitting} for details.}
    \label{fig:limbfit}
\end{figure*}

We considered two configurations of the positive chords for the limb fitting. For the first configuration, we kept the relative positions of the original chords but had to correct the absolute time of chord 8 (see Sect. \ref{ss:time_synch}). In this regard, we performed a weighted\footnote{The weights of the chords' centers were calculated as the inverse of $\frac{1}{2}\sqrt{\sigma_{ingress}^2+\sigma_{egress}^2}$.} least-squares polynomial fit of degree one to the centers of the chords in Table \ref{t:occtimes}, not using chord 8 for that fit. We then shifted chord 8 until its center lay on said linear regression and thus performed the elliptical fitting to the chords' extremities (Fig. \ref{fig:limbfit}a). For the second configuration, we aligned the centers of all the chords using the aforementioned linear regression and then performed the elliptical limb fitting to the chords' extremities (see Fig. \ref{fig:limbfit}b).

The alignment of the centers is a condition for the parallel chords of an ellipse, and the needed shifts are not within the centers' uncertainties. Although synchronization via NTP servers provides theoretical accuracies of \unit[0.01]{\si{\second}}, uncertainties of up to tenths of a second have been reported to arise from using different operating systems \citep{Barry2015}, camera software, or even due to delays in the shutter opening. However, adding a nominal average error to all the extremities would be unwise since such error should be the same for the ingress and egress of each chord, and we would be overestimating the error. On the other hand, the necessity of correcting unexplained time shifts has also been reported \citep{Elliot2010, BragaRibas2013}. So by aligning the chords, we account for possible systematic timing errors and small topography that could have decentered the chords.

We obtained the best elliptical fit to the extremities of the chords via minimization of the sum of squared residuals $\sum_i(d_{i,\mathrm{obs}} - d_{i,\mathrm{fit}})^2$, being $d_i$ the shortest distance between the $i$-extremity and the evaluated ellipse following the direction of the chord. The uncertainties of the elliptical parameters were determined using a Monte Carlo method. To do this, we generated 10\,000 random sets of extremities of the chords for each configuration, sampled from within the uncertainty bars, and then searched for the ellipse that minimizes the aforementioned equation. The Monte Carlo distributions obtained for each of the ellipse parameters are included in Appendix \ref{appendix_MC}. In Table \ref{t:param_elipse} we show the nominal value and the standard deviation of each distribution.

\begin{table*}[hbt!]
\centering
\caption{Results of the elliptical limb fitting.}
\label{t:param_elipse}
\begin{tabular}{l|cc}
\hline
Parameter                           & Original chords\tablefootmark{a}  & Aligned chords\\\hline\hline
Semi-major axis \textit{a'} (km)    & 292~$\pm$~3                       & 293~$\pm$~3\\
Semi-minor axis \textit{b'} (km)    & 231~$\pm$~6                       & 230~$\pm$~6\\
Oblateness $\epsilon'=(a'-b')/a'$   & 0.21~$\pm$~0.03                   & 0.21~$\pm$~0.03\\
Tilt angle (\degree)                & -11~$\pm$~2                       & -6~$\pm$~2 \\
($f_\mathrm{c}$, $g_\mathrm{c}$) (km)\tablefootmark{b}   & (-1010~$\pm$~2, 770~$\pm$~5)         & (-1007~$\pm$~2, 769~$\pm$~5)\\
($f_\mathrm{c}$, $g_\mathrm{c}$) (mas)\tablefootmark{b}  & (-38.57~$\pm$~0.09, 29.40~$\pm$~0.19)    & (-38.43~$\pm$~0.09, 29.37~$\pm$~0.17)\\
Area-equivalent diameter (km)       & 519~$\pm$~10                     & 519~$\pm$~10 \\
Mean area-equivalent diameter (km)  & 545~$\pm$~13                     & 545~$\pm$~13\\
Geometric albedo $p_\mathrm{V}$     & 0.134~$\pm$~0.010                     & 0.134~$\pm$~0.010 \\
\hline
\end{tabular}
\tablefoot{The nominal values are obtained by minimization of a least squares function and the uncertainties are the standard deviation of the Monte Carlo distributions, see Sect. \ref{ss:limb_fitting}. The corresponding distributions can be seen in appendix \ref{appendix_MC}. The uncertainty bars of the equivalent diameters and albedo are obtained analytically from their respective mathematical expressions.\\
\tablefoottext{a}{Chord 8 is the only chord that has been shifted for this case, see Sect. \ref{ss:limb_fitting}.}
\tablefoottext{b}{Offsets with respect to the JPL \#33 orbit.}}
\end{table*}

The instantaneous area-equivalent diameter of 2003~VS$_2$ is also given in Table \ref{t:param_elipse}. Given that the rotational phase of 2003~VS$_2$ was near its brightness minimum during the stellar occultation (see Fig. \ref{fig:rotationalLC}), the projected area was also near its minimum, and thus this diameter is a lower limit for the real equivalent diameter of 2003~VS$_2$. If we take the rotational phase and the rotational light curve amplitude into account, we can derive the mean area-equivalent diameter as:
\begin{equation}
\label{eq:area_eq_diameter}
    D_{\mathrm{mean}}=D_{\mathrm{occ}}\times 10^{(m_{occ}-m_{mean})/5},
\end{equation}
where $m_{occ}$ is the relative magnitude of 2003~VS$_2$ during the stellar occultation ($m_{occ}~=~0.104~\pm~0.010$ and $m_{mean}~=~0$, see Fig. \ref{fig:rotationalLC}), and $D_{occ}$ is its instantaneous area-equivalent diameter.

The derived mean area-equivalent diameter of \unit[545~$\pm$~13]{\si{\kilo\meter}} is slightly greater than the value obtained from Herschel radiometric data \citep[$D=523^{+35.1}_{-34.4}\si{\kilo\meter}$,][]{Mommert2012} but in agreement within the error bars. It is, however, slightly smaller than the area-equivalent diameter of \unit[553$^{+36}_{-33}$]{\si{\kilo\meter}} calculated with the data from \citet{BenedettiRossi2019} and applying Eq. \ref{eq:area_eq_diameter} to take into account the rotational phase at the occultation time, but still compatible within the uncertainty.

\subsection{Geometric albedo}

The geometric albedo at $V$-band of 2003~VS$_2$ at the time of the stellar occultation can be derived from:
\begin{equation}
    p_V = 10^{0.4(m_{V,\mathrm{Sun}}-H_{V,\mathrm{occ}})}/(A/\pi)
\end{equation}
being $m_\mathrm{V,Sun}$ the $V$-magnitude of the Sun ($m_\mathrm{V,Sun}$=-26.74 mag), $H_\mathrm{V,occ}$ the instantaneous absolute magnitude of 2003~VS$_2$ in $V$-band at the time of the stellar occultation, and \textit{A} the projected area of 2003~VS$_2$ during the event in astronomical units squared.

To obtain $H_\mathrm{V,occ}$ we corrected 2003~VS$_2$'s rotationally averaged $H_{\rm V}$ (\unit[4.14 $\pm$ 0.07]{mag}; \citet{AlvarezCandal2016} and priv. comm.) by adding the theoretical relative magnitude of 2003~VS$_2$ during the stellar occultation, which is given by the Fourier fit to the rotational light curve and has a value of 0.104~$\pm$~0.010 mag at the $1\sigma$-level.

The obtained geometric albedo of 0.134~$\pm$~0.010 (Table \ref{t:param_elipse}) is slightly smaller but compatible with the one derived from the combination of Herschel thermal measurements and Spitzer data \citep[$0.147^{+0.063}_{-0.043}$,][]{Lellouch2013}, and is also in agreement with the last derived albedo from \citet{BenedettiRossi2019} (0.131~$^{+0.024}_{-0.013}$), see Table \ref{t:vs}. 

If we use the same absolute magnitude as \citet{Lellouch2013} ($H_{\rm V} =$ \unit[4.11 $\pm$ 0.38]{mag}) to calculate the albedo, it has a value of 0.14~$\pm$~0.05, which is still smaller but compatible within the error bars.

\subsection{Size and shape}
\label{ss:size and shape}

The double-peaked rotational light curve of 2003~VS$_2$ suggests that it is either a rotating triaxial ellipsoid or an oblate spheroid with a significant irregularity or a large albedo variation along its surface; this second case is unlikely, especially considering 2003~VS$_2$'s large light curve amplitude. Besides, the two maxima and minima per rotation cycle are different, which usually indicates a triaxial shape with some albedo spots or small topographic features, since we do not expect a perfectly symmetrical and homogeneous body orbiting in space. This is the case for many observed TNOs with large rotational light curve amplitudes, such as Haumea \citep{Lacerda2010}, Varuna \citep{Fernandez-Valenzuela2019}, and 2008~OG$_{19}$ \citep{FernandezValenzuela2016}. On the other hand, if not triaxial, the rotation period of 2003~VS$_2$ would be \unit[3.6]{\si{\hour}}, which is probably too fast for a TNO (no TNO is known to rotate that fast). Hence, assuming a triaxial ellipsoidal shape for 2003~VS$_2$, we searched for the axes $a>b>c$ of a triaxial body, rotating around its $c$-axis, that would give the observed elliptical projection during the stellar occultation (Fig. \ref{fig:limbfit}) while also showing an amplitude of $\Delta m$~=~\unit[0.264~$\pm$~0.017]{mag} on its rotational light curve.

To do this, we used the procedures described in \citet{GendzwillStauffer1981} to generate ellipsoids characterized by three semi-major axes and three orientation angles with respect to the Cartesian system, and then project these ellipsoids into a plane perpendicular to the line of sight, in order to compare that projection with the instantaneous limb of 2003~VS$_2$ during the stellar occultation (Fig. \ref{fig:limbfit}).

An additional constraint to the model is the observed rotational light curve amplitude of 2003~VS$_2$ (Fig. \ref{fig:rotationalLC}). To include it, we implemented the well-known relation between the peak-to-peak variation of the rotational light curve of a small body and its three principal semi-axes \citep{Binzel1989}:
\begin{equation}
\label{eq:rlc_ampl_asteroid}
    \Delta m = 2.5 \log \left( \frac{a}{b}\right) - 1.25 \log \left(\frac{a^2 \cos^2(\theta)+c^2\sin^2(\theta)}{b^2 \cos^2(\theta)+c^2\sin^2(\theta)}\right) 
\end{equation}
where $\theta$ is the polar aspect viewing angle, that is, the angle between the rotation axis (c, in this case) and the line of sight.

For each of the ellipses obtained via Monte Carlo for the limb-fitting (Sect. \ref{ss:limb_fitting}, Fig. \ref{fig:limbfit}), we searched for the ellipsoid that would give the most similar projection in terms of least squares minimization. Although the limb fit obtained in Sect. \ref{ss:limb_fitting} has virtually the same values for both considered chord configurations, we decided to search for the corresponding ellipsoid of both solutions to see if the slight difference in the tilt angle would give different 3D shapes. The obtained distributions for the ellipsoid's semi-axes, aspect angle, and the corresponding rotational light curve amplitude derived from Eq. \ref{eq:rlc_ampl_asteroid} are shown in appendix \ref{appendix_MC}. In Table \ref{t:param_ellipsoid} we show the best 3D fit (which is the one corresponding to the best limb fit in Table \ref{t:param_elipse}), and the uncertainty bars are given as the standard deviation of the Monte Carlo distributions. According to the results, the obtained triaxial ellipsoid would produce a variation of $~$0.24 mag during its rotation so, as stated before, the remaining observed rotational light curve amplitude might be due to albedo spots or topographic features that cannot be studied with the available data; we would either need more chords from the stellar occultation to study topographic features, or rotational light curves obtained with different filters to study albedo variations. The differences in the rotational light curve amplitude due to albedo spots would be on the order of a few cents of a magnitude (this is the range of variability due to albedo variegations observed in large TNOs such as Varuna, Haumea, 2008~OG$_{19}$) Considering this and given that a model of a triaxial body without albedo spots to explain the light curve amplitude is a simplification, we also computed the ellipsoid accepting solutions that give a rotational light curve amplitude 0.06 mag less than the observed value, to allow for large albedo spots. These results are also included in Table \ref{t:param_ellipsoid} and appendix \ref{appendix_MC}. Finally, the spherical volume equivalent diameter ($D_\mathrm{V_{eq}}$) derived from these parameters is also presented in Table \ref{t:param_ellipsoid}.

\begin{table*}[]
    \centering
    \caption{Results of the 3D-fit of 2003~VS$_2$.}
    \begin{tabular}{l|cc|cc}
    \hline
    \multirow{2}{*}{Parameter}& \multicolumn{2}{c|}{Observed $\Delta m$} & \multicolumn{2}{c}{$\Delta m_{min}$ = 0.18 mag}\\ 
                                       & Original chords\tablefootmark{a}  & Shifted chord & Original chords\tablefootmark{a}  & Shifted chords\\ \hline\hline
         Semiaxis \textit{a} (km)               & 339~$\pm$~5               & 338~$\pm$~5           & 327~$\pm$~5               & 330~$\pm$~5\\
         Semiaxis \textit{b} (km)               & 235~$\pm$~6               & 233~$\pm$~7           & 233~$\pm$~6               & 235~$\pm$~9\\
         Semiaxis \textit{c} (km)               & 226~$\pm$~8               & 227~$\pm$~7           & 228~$\pm$~8               & 222~$\pm$~8\\
         Aspect angle $\theta$ (\degree)        & 59~$\pm$~2                & 57~$\pm$~3            & 51~$\pm$~2               & 51~$\pm$~6\\
         Rotational light curve amplitude (mag) & 0.2497~$\pm$~0.0008       & 0.2489~$\pm$~0.0009   & 0.187~$\pm$~0.003         & 0.184~$\pm$~0.003\\
         $a/b$                                  & 1.44~$\pm$~0.06           & 1.45~$\pm$~0.06       & 1.41~$\pm$~0.06               & 1.40~$\pm$~0.07\\
         $b/c$                                  & 1.040~$\pm$~0.017         & 1.022~$\pm$~0.018     & 1.02~$\pm$~0.02               & 1.06~$\pm$~0.03\\
         Spherical volume equivalent diameter (km)   & 524~$\pm$~7          & 523~$\pm$~7           & 517~$\pm$~7               & 516~$\pm$~8\\
         \hline

    \end{tabular}
    \tablefoot{We present the ellipsoids obtained using the observed rotational light curve amplitude ($\Delta m$~=~\unit[0.264~$\pm$~0.017]{mag}) and the ones allowing for bigger albedo spots, see Sect. \ref{ss:size and shape}. The nominal values of the parameters are the ones obtained from the optimal ellipses in Table \ref{t:param_elipse}. The uncertainties are given as the standard deviation of the Monte Carlo distributions shown in appendix \ref{appendix_MC}, for the two chords' configurations considered for the limb-fitting.}\\
    \tablefoottext{a}{Chord 8 is the only chord that has been shifted for this case, see Sect. \ref{ss:limb_fitting}.}
    \label{t:param_ellipsoid}
\end{table*}

Both obtained solutions are almost compatible with the one presented in \citet{BenedettiRossi2019} within the errors. The differences might arise from the fact that, in the previous work, the calculations were simplified by considering that 2003~VS$_2$ was on its brightness maximum during the 2014 occultation, when the actual rotational phase at the time of the occultation was +0.07 with respect to the brightness maximum. Moreover, the rotational light curve amplitude they used was smaller than the one used in this work. However, we checked that the two amplitudes cannot be explained simultaneously by a change of the aspect angle.

\begin{figure}
    \centering
    \resizebox{\hsize}{!}{\includegraphics{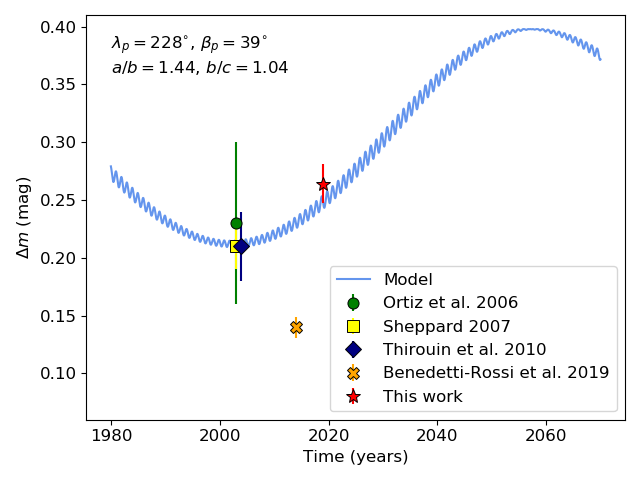}}
    \caption{Theoretical long-term variability of 2003~VS$_2$'s rotational light curve amplitude and observational results found in the literature.}
    \label{fig:m_variability}
\end{figure}

We derived the ecliptic coordinates of the pole $(\lambda_p, \beta_p)$ by simultaneously minimizing the difference between the observed rotational light curve amplitude and the one derived using Eq. \ref{eq:rlc_ampl_asteroid}, and the difference between the aspect angle $\delta$ derived in Sect. \ref{ss:size and shape} and the one derived via:
\begin{equation}
    \delta = \frac{\pi}{2} - \arcsin\left[\sin{\beta_e}\sin{\beta_p} +  \cos{\beta_e}\cos{\beta_p}\cos{(\lambda_e - \lambda_p)} \right]
\end{equation}
being $(\lambda_e, \beta_e)$ the ecliptic coordinates of the sub-Earth point in the object-centered reference frame at the time of this work's stellar occultation. The pole's ecliptic coordinates $(\lambda_p, \beta_p)$ derived are (228\degree, 39\degree) (for a more detailed explanation see \citet{FernandezValenzuela2022} and references therein). Combining the pole's coordinates, 2003~VS$_2$'s ephemeris, and the three-dimensional axes obtained in Sect. \ref{ss:size and shape}, we obtained the theoretical variation of the rotational light curve amplitude through the years and compared it to the values in the literature. The results are plotted in Fig. \ref{fig:m_variability}. It can be seen that the previously published results were obtained during the minimum of the rotational light curve amplitude, but it has been increasing since 2005. The published value found in \cite{BenedettiRossi2019} is much lower than theoretically expected, but the remaining measurements do agree with the model. We suspect that the rotational light curve in \citet{BenedettiRossi2019} may have been contaminated by a background star or had some yet unidentified technical problem or reduction artifact because there are no physical scenarios that can explain a sudden decrease in the amplitude except perhaps a sudden brightening due to dust release through sublimation activity or through a collision, but this is very unlikely.

The projection of the ellipsoids obtained from the original-chords configuration at the rotational phase in which the stellar occultation in \citet{BenedettiRossi2019} happened (+0.07 with respect to the brightness maximum) gives ellipses with semi-axes $a~\times~ b = 305\pm4~\times~ 230\pm6$ km (for the observed rotational light curve amplitude) and $a~\times~ b = 302\pm4~\times~ 230\pm6$ km (for a minimum rotational light curve amplitude of 0.18 mag), which are in agreement with the limb fit reported in \citet{BenedettiRossi2019}.

\begin{figure}
    \resizebox{\hsize}{!}{\includegraphics{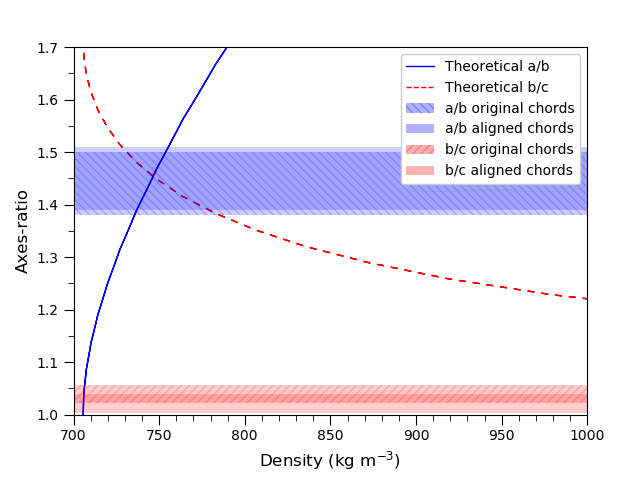}}
    \caption{Theoretical axes-ratios $a/b$ (solid blue line) and $b/c$ (dashed red line) of a three-dimensional body in hydrostatic equilibrium, rotating with a period of \unit[7.41753 $\pm$ 0.00001]{\si{\hour}}, for different densities \citep{Chandrasekhar1987}. The axes-ratios from the observed data are plotted as colored bands, for the both considered cases. The theoretical $b/c$ ratios from hydrostatic equilibrium do not agree with the observations for any possible density of a homogeneous body because there is no intersection of the dashed red line with the pink band.}
    \label{fig:ratios_density}
\end{figure}

To conclude, we can compare these results with the theoretical axes-ratios of a triaxial ellipsoid with the rotation period of 2003~VS$_2$ and a homogeneous density, to see if the derived shape is compatible with a hydrostatic-equilibrium figure. Using the formalism from \citet{Chandrasekhar1987}, we searched for the theoretical values of the axes-ratios \textit{a/b} and \textit{b/c} (solid and dashed lines in Fig. \ref{fig:ratios_density}, respectively) of a triaxial ellipsoid with a rotation period of \unit[7.41753 $\pm$ 0.00001]{\si{\hour}} and homogeneous densities between \unit[600 and 1000]{\si{\kilo\gram\per\cubic\meter}}. We also plotted in Fig. \ref{fig:ratios_density} the axes-ratios bands derived from the stellar occultation (their values are presented in Table \ref{t:param_ellipsoid}). In the figure, one can see that there is an overlapping region between the theoretical and observed axes-ratio \textit{a/b}, but that is not the case for the axes-ratio \textit{b/c}. Therefore, we can conclude that the derived shape of 2003~VS$_2$ is not consistent with the hydrostatic-equilibrium figure of a homogeneous body with the rotation period of 2003~VS$_2$ for any density value. This result suggests that, similar to the case of Haumea \citep{Ortiz2017}, we might need to consider granular physics to explain the body's shape, because in this case, a differentiated body is less plausible due to 2003~VS$_2$'s smaller size. Actually, the differentiation obtained by \citet{Loveless2022} for icy bodies with radii larger than 200~km, although not negligible, is so small that it would probably not produce a significant deviation from the equilibrium shape for a homogeneous body of the size of 2003~VS$_2$. However, that model excludes or simplifies some of the physical and chemical processes and considers pure spherical bodies, so we cannot discard some differentiation. The possibility that 2003~VS$_2$ might be sustaining some stress seems more plausible but, in reality, we can only speculate that the deviation of its shape from that of pure hydrostatic equilibrium could be due to one or a combination of both scenarios.

\section{Conclusions}

On 2019 October 22, we observed the stellar occultation of the GAIA source 3449076721168026496 (m$_V$~=~14.1 mag) caused by the plutino (84922) 2003~VS$_2$. Out of the 39 participant observing sites, 12 reported a positive detection, located in Bulgaria, Romania, and Serbia. Two positive chords were combined into a single one, giving 11 effective positive chords. This is one of the best observed stellar occultations by a TNO so far.

The projected shape of 2003~VS$_2$ was fitted to an ellipse considering two configurations of the positive chords: for the first one, we fitted the extremities of the original chords obtained from the stellar occultation after shifting only one chord due to a problem with its absolute time; for the second case we shifted all the positive chords to align their centers. The best solution for the instantaneous limb of 2003~VS$_2$ was obtained by minimization of the sum of squared residuals, and the uncertainties were derived via a Monte Carlo method and correspond to the standard deviation of the obtained distributions. The result has virtually the same value for both configurations as can be seen in Table \ref{t:param_elipse}; for the original configuration it has semi-axes \textit{a'$_1$}~=~\unit[292~$\pm$~3]{\si{\kilo\meter}} and \textit{b'$_1$}~=~\unit[231~$\pm$~6]{\si{\kilo\meter}}, and a tilt angle of $-$11\degree~$\pm$~2\degree.

We carried out photometric observations of 2003~VS$_2$ two days after the stellar occultation and derived an amplitude of $\Delta m$~=~\unit[0.264~$\pm$~0.017]{mag} from the obtained rotational light curve. This value is slightly greater than the ones previously published for this object, but can be explained through a change in the aspect angle of the body. Taking this and the rotational phase during the stellar occultation into account, the mean area-equivalent diameter of 2003~VS$_2$ is $D_{\mathrm{A_{eq}}}$~=~\unit[545~$\pm$~13]{\si{\kilo\meter}}. The derived geometric albedo during the stellar occultation is $p_\mathrm{V}$~=~0.134~$\pm$~0.010. These values are in agreement with the ones published in \citet{BenedettiRossi2019}, and with the diameter derived from thermal models and the radiometric albedo obtained with Herschel and Spitzer data within the error bars \citep[$D=523^{+35.1}_{-34.4}$km and $p_\mathrm{V}$~=~0.147$^{+0.063}_{-0.043}$,][]{Mommert2012}.

We derived the 3D shape of 2003~VS$_2$ by combining its rotational light curve and the instantaneous projection and assuming a triaxial ellipsoidal shape. The obtained ellipsoid has principal semi-axes \textit{a}~=~\unit[339~$\pm$~5]{\si{\kilo\meter}}, \textit{b}~=~\unit[235~$\pm$~6]{\si{\kilo\meter}}, and \textit{c}~=~\unit[226~$\pm$~8]{\si{\kilo\meter}}, with an aspect angle of either $\theta_1$~=~59\degree~$\pm$~2\degree or its supplementary $\theta_2$~=~121\degree~$\pm$~2\degree, depending on the considered sense of rotation. If we allow for a 0.1 mag variability due to albedo spots in the 3D model, the resulting ellipsoid has a semi-major axis $\sim$~10~km smaller and fully compatible with the projected shape in \citet{BenedettiRossi2019}. These values give a spherical-volume equivalent diameter of $D_\mathrm{V_{eq}}$~=~\unit[524~$\pm$~7]{\si{\kilo\meter}}. This solution is not compatible with a homogeneous body in hydrostatic equilibrium rotating with the known period of 2003~VS$_2$, requiring differentiation or, most likely, an internal structure that can sustain stress to some degree.

We found no evidence of a dense ring or debris material orbiting around 2003~VS$_2$, of the type seen in Chariklo \citep{BragaRibas2014}.

\begin{acknowledgements}
We acknowledge financial support from the State Agency for Research of the Spanish MCIU through the "Center of Excellence Severo Ochoa" award to the Instituto de Astrof\'isica de Andaluc\'ia (SEV-2017-0709). Funding from Spanish projects PID2020-112789GB-I00 from AEI and Proyecto de Excelencia de la Junta de Andaluc\'ia PY20-01309 is acknowledged. Part of the research leading to these results has received funding from the European Research Council under the European Community’s H2020 (2014-2020/ERC Grant Agreement no. 669416 “LUCKY STAR”).
M.V-L. acknowledges funding from Spanish project AYA2017-89637-R (FEDER/MICINN). P.S-S. acknowledges financial support by the Spanish grant AYA-RTI2018-098657-J-I00 ``LEO-SBNAF''.
Part of the work of M.P. was financed by a grant of the Romanian National Authority for Scientific Research and Innovation, CNCS - UEFISCDI, PN-III-P1-1.1-TE-2019-1504. E.F.-V. acknowledges financial support from the Florida Space Institute and the Space Research Initiative.
The following authors acknowledge the respective CNPq grants: F.B-R 309578/2017-5; B.E.M. 150612/2020-6; RV-M 304544/2017-5, 401903/2016-8; J.I.B.C. 308150/2016-3 and 305917/2019-6; M.A 427700/2018-3, 310683/2017-3 and 473002/2013-2.
D.I. and O.V. acknowledge funding provided by the Ministry of Education, Science, and Technological Development of the Republic of Serbia (contracts 451-03-9/2021-14/200104, 451-03-9/2021-14/200002). D.I. acknowledges the support of the Alexander von Humboldt Foundation.
M.H. thanks the Slovak Academy of Sciences (VEGA No. 2/0059/22) and the Slovak Research and Development Agency under the Contract No. APVV-19-0072. This work has also been supported by the VEGA grant of the Slovak Academy of Sciences No. 2/0031/18. A.P., R.S. and C.K. acknowledge the grant of K-138962 of National Research, Development and Innovation Office (Hungary).
This study was financed in part by the Coordenação de Aperfeiçoamento de Pessoal de Nível Superior - Brasil (CAPES) - Finance Code 001 and the National Institute of Science and Technology of the e-Universe project (INCT do e-Universo, CNPq grant 465376/2014-2). This work has made use of data from the European Space Agency (ESA) mission Gaia (\url{https://www.cosmos.esa.int/gaia)}, processed by the Gaia Data Processing and Analysis Consortium (DPAC, \url{https://www.cosmos.esa.int/web/gaia/dpac/consortium)}. Funding for the DPAC has been provided by national institutions, in particular the institutions participating in the Gaia Multilateral Agreement. This research is partially based on observations collected at the Centro Astron\'omico Hispano-Alem\'an (CAHA) at Calar Alto, operated jointly by Junta de Andaluc\'ia and Consejo Superior de Investigaciones Cient\'ificas (IAA-CSIC). This research is also partially based on observations carried out at the Observatorio de Sierra Nevada (OSN) operated by Instituto de Astrof\'isica de Andaluc\'ia (CSIC). This article is also based on observations made in the Observatorios de Canarias del IAC with the Liverpool Telescope operated on the island of La Palma by the Instituto de Astrof\'isica de Canarias in the Roque de los Muchachos Observatory.
\end{acknowledgements}

\bibliographystyle{aa}
\bibliography{2003VS2-2019Oct22-bibtex}

\begin{appendix}

\section{Observing sites involved in the occultation campaign that could not observe}
\label{appendix_observers}

In this appendix we include the list of contacted observing sites that could not observe due to bad weather or technical problems.

\begin{table}[!htbp]
    \centering
    \caption{Details of the observing sites involved in the 2003~VS$_2$ 2019 October 22 occultation that could not observe.}
    \begin{tabular}{*5{l}}
        \hline
        Site name   & Latitude (N)  & Telescope  &              &  \\
        Location    & Longitude (E) & aperture &  Observers   & Problem   \\
        MPC Code    &               & (cm)                   &              &           \\
        \hline\hline
        
        Gala\c ti Observatory ($\times$2)   & 45\degr~ 25\arcmin~ 7\farcs9  & 40 & O. Tercu              &\\
        Romania                 & 28\degr~ 1\arcmin~ 57\arcsec    & 20 & A. M. Stoian          & Bad Weather\\
        C73                     &               & & G. Neagu      &\\
        &&& D.Zlat &\\\hline
        
        Astroclubul Bucure\c sti ($\times$2)  & --    & 40 & D. Berte\c{s}teanu&\\
        Romania (mobile telescope)   &               & 20   & M. H. Naiman & Technical problems \\
        --          &               &             & Z. Deak&\\\hline
        
                    & --    & &&\\
        Romania (mobile telescope)    &               &20& B. Dumitru & Technical problems\\
        --          &               && &\\\hline
        
        B\^arlad Observatory    & 46\degr~ 13\arcmin~ 54\farcs1    &&                  & \\
        Romania                 & 27\degr~ 40\arcmin~ 10\farcs2    &20& C. Vantdevara    & Bad weather\\
        L22                     &               &&                  &\\\hline
        
        G.V.Schiaparelli Observatory & 45\degr~ 52\arcmin~04\arcsec  &&&\\
        Varese, Italy                & $-$08\degr~ 46\arcmin~15\arcsec &84& L. Buzzi & Bad weather \\
        204 &&&&\\\hline
        
        San Marcello Pistoiese & 44\degr~ 3\arcmin~ 46\farcs9 &&&\\
        Italy & 10\degr~ 48\arcmin~ 15\farcs1 &60 & P. Bacci & Technical problems\\
        104 &&&&\\\hline
        
        ISON Uzhgorod Observatory & 48\degr~ 33\arcmin~ 48\farcs2 &&&\\
        Derenivka, Ukraine & 22\degr~ 27\arcmin~ 12\farcs6 &40& V. Kudak & Technical problems\\
        K99 &&&&\\\hline
        
        Odessa-Mayaki & 46\degr~ 23\arcmin~ 50\farcs2   && Y. Krugly &\\
        Ukraine & 30\degr~ 16\arcmin~ 18\farcs1         &80& I. Belskaya & Bad weather \\
        583 &                       &&&\\\hline
        
        Zalistci  & 48\degr~ 50\arcmin~ 53\farcs9  && Y. Krugly &\\
        Ukraine   & 26\degr~ 43\arcmin~ 5\farcs9    &30& I. Belskaya & Bad weather \\
        L18       &                 &&           &\\\hline
        
        Uzhgorod-Derenivka, & 48\degr~ 33\arcmin~ 48\farcs2 && Y. Krugly&\\
        Ukraine & 22\degr~ 27\arcmin~ 12\farcs599 &40& I. Belskaya & Bad weather\\
        K99 &&&&\\ \hline
        
        Kyiv-Comet Station & 50\degr~ 17\arcmin~ 52\farcs8 && Y. Krugly &\\
        Ukraine & 30\degr~ 31\arcmin~ 28\farcs6 &70& I. Belskaya & Bad weather \\
        585 &&&&\\\hline
        
        Kharkiv-Chuguev Station & 49\degr~ 38\arcmin~ 34\farcs6 && Y. Krugly &\\
        Ukraine & 36\degr~ 56\arcmin~ 12\farcs8 & 70& I. Belskaya & Bad weather \\
        121 &&&&\\\hline
        
        Hvar Observatory & 43\degr~ 10\arcmin~42\arcsec && D. Ruzdjak &\\
        Hvar, Croatia & 16\degr~ 26\arcmin~54\arcsec &106& S. Cikota & Bad weather \\
        -- &&&&\\\hline
    \end{tabular}
    \label{t:noobs_sites}
\end{table}   

\FloatBarrier

\section{Monte Carlo distributions}
\label{appendix_MC}
In this appendix we present the obtained Monte Carlo distributions for the limb-fitting and the 3D-fitting of 2003~VS$_2$, respectively discussed in Sects. \ref{ss:limb_fitting} and \ref{ss:size and shape}, for the two considered distributions of the positive chords.

\begin{figure*}[!htb]
    \centering
    \includegraphics[width=17cm]{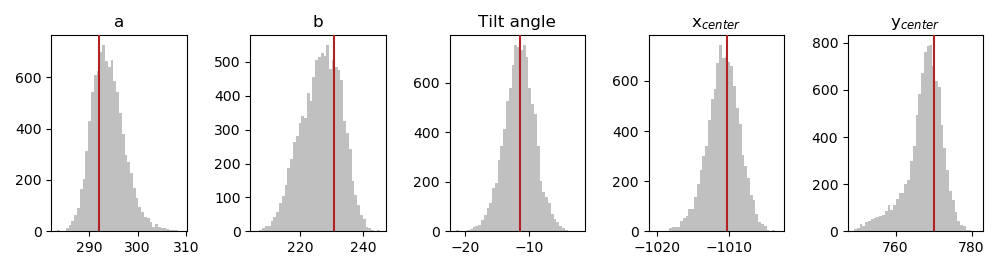}
    \caption{Monte Carlo distributions for the semi-axes a and b (in \si{\kilo\meter}), the tilt angle (in degrees) and the coordinates (x,y) of the center of the ellipse, from the elliptical fit to the unshifted chords (with only chord 8 shifted) in Fig. \ref{fig:limbfit}a, see Sect. \ref{ss:limb_fitting}. The vertical red lines show the value of the parameters for the best elliptical fit via minimization of the sum of squared residuals, see Table \ref{t:param_elipse}.}
    \label{fig:limbfithist_sind}
\end{figure*}

\begin{figure*}[!t]
    \centering
    \includegraphics[width=17cm]{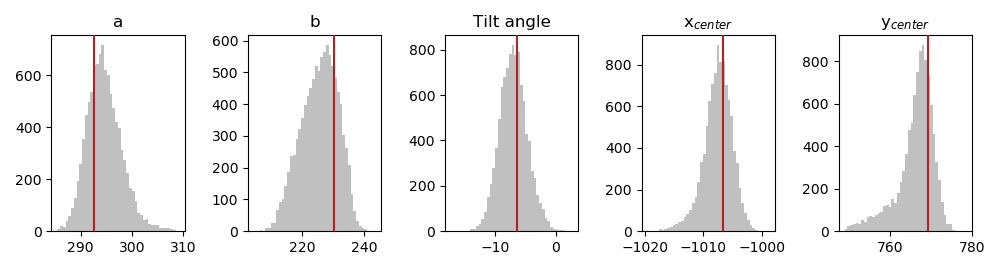}
    \caption{Monte Carlo distributions for the semi-axes a and b (in \si{\kilo\meter}), the tilt angle (in degrees) and the coordinates (x,y) of the center of the ellipse, from the elliptical fit to the aligned chords in Fig. \ref{fig:limbfit}b, see Sect. \ref{ss:limb_fitting}. The vertical red lines show the value of the parameters for the best elliptical fit via minimization of the sum of squared residuals, see Table \ref{t:param_elipse}.}
    \label{fig:limbfithist_d}
\end{figure*}

\begin{figure*}[!t]
    \centering
    \includegraphics[width=17cm]{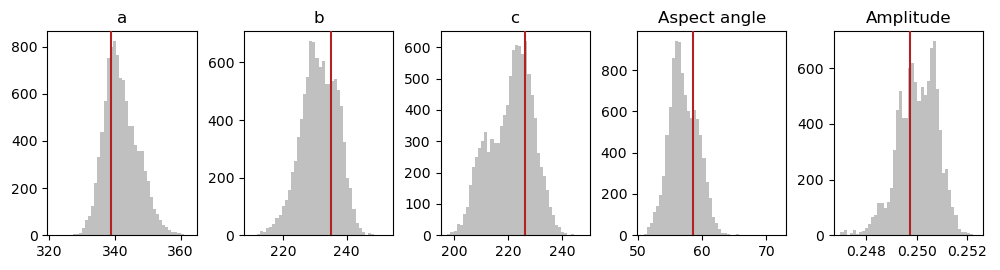}
    \caption{Monte Carlo distributions for the semi-axes a, b, and c  (in \si{\kilo\meter}) of a triaxial ellipsoid compatible with the occultation observation and observed rotational light curve amplitude, as well as distributions of the aspect angle and derived rotational light curve amplitude obtained via Eq. \ref{eq:rlc_ampl_asteroid}, for the case of the original chords with only chord 8 shifted, see Sects. \ref{ss:limb_fitting} and \ref{ss:size and shape}. The vertical red lines show the ellipsoidal values corresponding to the best elliptical fit, see Table \ref{t:param_ellipsoid}.}
    \label{fig:3dhist_sind}
\end{figure*}

\begin{figure*}[!t]
    \centering
    \includegraphics[width=17cm]{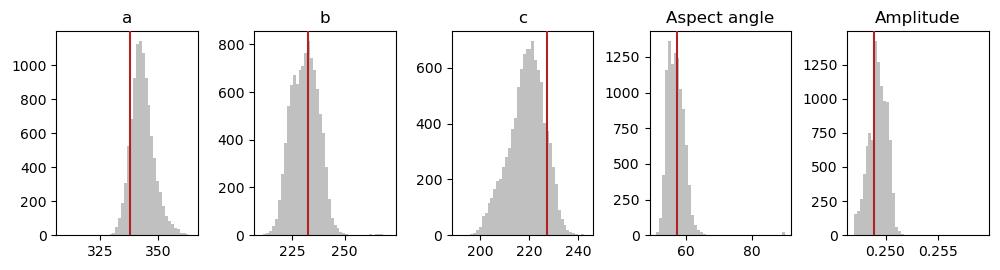}
    \caption{Monte Carlo distributions for the semi-axes a, b, and c (in \si{\kilo\meter}) of a triaxial ellipsoid compatible with the occultation observation and observed rotational light curve of 2003~VS$_2$, as well as distributions of the aspect angle and derived rotational light curve amplitude obtained via Eq. \ref{eq:rlc_ampl_asteroid}, for the case of all the chords shifted, see Sects. \ref{ss:limb_fitting} and \ref{ss:size and shape}. The vertical red lines show the ellipsoidal values corresponding to the best elliptical fit, see Table \ref{t:param_ellipsoid}.}
    \label{fig:3dhist_d}
\end{figure*}

\begin{figure*}[t!]
    \centering
    \begin{subfigure}[t]{\columnwidth}
    \centering
        \includegraphics[width=\columnwidth]{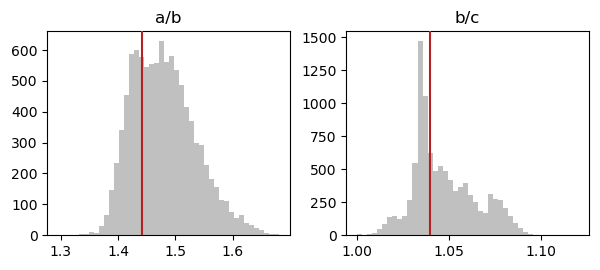}
        \caption{}
        \label{subfig:ratios_ejes_sind}
    \end{subfigure}
    ~
    \begin{subfigure}[t]{\columnwidth}
    \centering
        \includegraphics[width=\columnwidth]{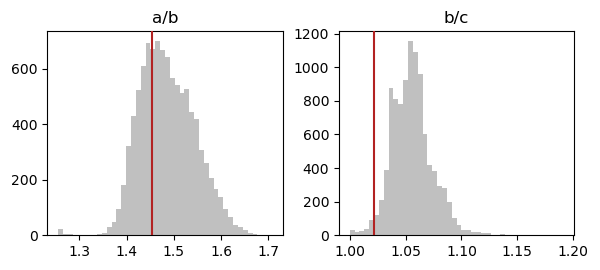}
        \caption{}
        \label{subfig:ratios_ejes_d}
    \end{subfigure}
    \caption{Monte Carlo distributions of the axes-ratios $a/b$ and $b/c$ of the fitted ellipsoid for the considered case of (a) the chords unshifted with only chord 8 shifted, and (b) the aligned chords, for the observed rotational light curve amplitude. The vertical red lines show the axes-ratios derived from the best elliptical fit, see Table \ref{t:param_ellipsoid}.}
    \label{fig:ratios_ejes}
\end{figure*}

\begin{figure*}[!t]
    \centering
    \includegraphics[width=17cm]{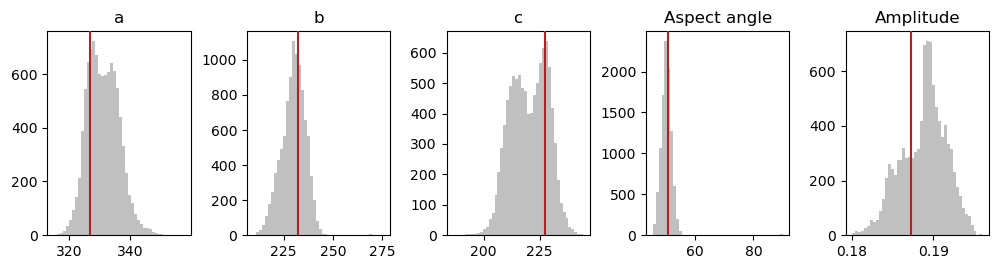}
    \caption{Monte Carlo distributions for the semi-axes a, b, and c  (in \si{\kilo\meter}) of a triaxial ellipsoid compatible with the occultation observation and allowing a minimum rotational light curve amplitude of 0.18 mag, as well as distributions of the aspect angle and derived rotational light curve amplitude obtained via Eq. \ref{eq:rlc_ampl_asteroid}, for the case of the original chords with only chord 8 shifted, see Sects. \ref{ss:limb_fitting} and \ref{ss:size and shape}. The vertical red lines show the ellipsoidal values corresponding to the best elliptical fit, see Table \ref{t:param_ellipsoid}.}
    \label{fig:3dhist_sind_m018}
\end{figure*}

\begin{figure*}[!t]
    \centering
    \includegraphics[width=17cm]{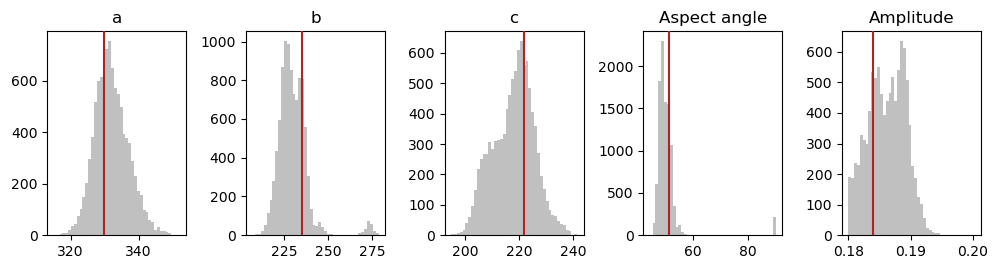}
    \caption{Monte Carlo distributions for the semi-axes a, b, and c (in \si{\kilo\meter}) of a triaxial ellipsoid compatible with the occultation observation and allowing a minimum rotational light curve amplitude of 0.18 mag, as well as distributions of the aspect angle and derived rotational light curve amplitude obtained via Eq. \ref{eq:rlc_ampl_asteroid}, for the case of all the chords shifted, see Sects. \ref{ss:limb_fitting} and \ref{ss:size and shape}. The vertical red lines show the ellipsoidal values corresponding to the best elliptical fit, see Table \ref{t:param_ellipsoid}.}
    \label{fig:3dhist_d_m018}
\end{figure*}

\begin{figure*}[t!]
    \centering
    \begin{subfigure}[t]{\columnwidth}
    \centering
        \includegraphics[width=\columnwidth]{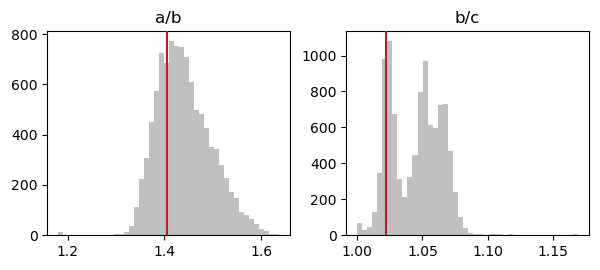}
        \caption{}
        \label{subfig:ratios_ejes_sind_m018}
    \end{subfigure}
    ~
    \begin{subfigure}[t]{\columnwidth}
    \centering
        \includegraphics[width=\columnwidth]{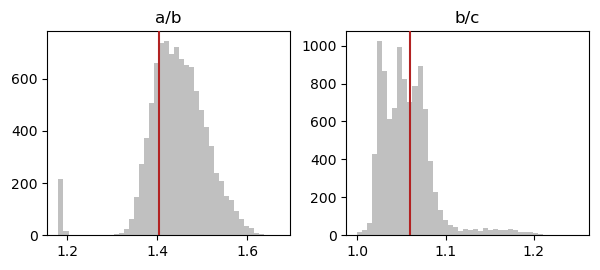}
        \caption{}
        \label{subfig:ratios_ejes_d_m018}
    \end{subfigure}
    \caption{Monte Carlo distributions of the axes-ratios $a/b$ and $b/c$ of the fitted ellipsoid for the considered case of (a) the chords unshifted with only chord 8 shifted, and (b) the aligned chords, if we allow a minimum rotational light curve amplitude of 0.18 mag. The vertical red lines show the axes-ratios derived from the best elliptical fit, see Table \ref{t:param_ellipsoid}.}
    \label{fig:ratios_ejes_m018}
\end{figure*}

\end{appendix}

\end{document}